\documentclass[iop,apj,tighten]{emulateapj}
\usepackage[breaklinks,colorlinks,urlcolor=blue,citecolor=blue,linkcolor=blue]{hyperref}
\usepackage{cleveref}
\usepackage{graphicx}

\usepackage{bm}
\usepackage{amsmath,amssymb}

\newcommand{\Msun}{{\rm M}_\odot}

\newcommand\lsim{\mathrel{\rlap{\lower4pt\hbox{\hskip1pt$\sim$}}
        \raise1pt\hbox{$<$}}}
\newcommand\gsim{\mathrel{\rlap{\lower4pt\hbox{\hskip1pt$\sim$}}
        \raise1pt\hbox{$>$}}}

\shorttitle{Growing by stellar bombardment}
\shortauthors{Tagawa et al.}

\begin{document}

\title{Making a supermassive star by stellar bombardment}

\author{Hiromichi Tagawa\altaffilmark{1}, Zolt{\'a}n Haiman\altaffilmark{2}, and Bence Kocsis\altaffilmark{1}}
\affil{\altaffilmark{1}Institute of Physics, E{\"o}tv{\"o}s University, P{\'a}zm{\'a}ny P.s., Budapest, 1117, Hungary\\
\altaffilmark{2}Department of Astronomy, Columbia University, 550 W. 120th St., New York, NY 10027, USA
}
\email{E-mail: htagawa@caesar.elte.hu}

\begin{abstract} 
Approximately two hundred supermassive black holes (SMBHs) have been discovered within the first $\sim$Gyr after the Big Bang. One pathway for the formation of SMBHs is through the collapse of supermassive stars (SMSs). A possible obstacle to this scenario is that the collapsing gas fragments and forms a cluster of main-sequence stars. Here we raise the
possibility that stellar collisions may be sufficiently frequent and energetic to inhibit the contraction of the massive protostar, avoiding strong UV radiation driven outflows, and allowing it to continue growing into an SMS. 
We investigate this scenario with semianalytic models incorporating star formation, gas accretion, dynamical friction from stars and gas, stellar collisions, and gas ejection.  We find that when the collapsing gas fragments at a density of $\lesssim 3\times 10^{10}\,\mathrm{cm^{-3}}$, the central protostar contracts due to infrequent stellar mergers, and in turn photoevaporates the remaining collapsing gas, resulting in the formation of a $\lesssim 10^4\,{\rm M_\odot}$ object.  On the other hand, when the collapsing gas fragments at higher densities (expected for a
metal-poor cloud with $Z\lesssim10^{-5}\,{\rm Z_\odot}$ with suppressed ${\rm H_2}$ abundance) the central protostar avoids contraction and keeps growing via frequent stellar mergers, reaching masses as high as $\sim 10^5-10^6\,{\rm M_\odot}$. We conclude that frequent stellar mergers represent a possible pathway to form massive BHs in the early universe.  
\end{abstract}

\section{Introduction}

Over the past decade, approximately two hundred supermassive black holes (SMBHs) have been discovered with masses of $\gtrsim 10^9\,{\rm M_\odot}$ at redshift $z\gtrsim6$ \citep[e.g.][and references therein]{Fan01,Willott10,Mortlock11,Venemans13,DeRosa14,Wu15,Jiang16,Matsuoka18,Banados18,YangJ19,WangR19,Onoue19,Izumi19,Shen19,Matsuoka+2019}. The formation process of these SMBHs remains one of the most puzzling problems in astrophysics \citep[see, e.g.][for reviews]{Volonteri10,Haiman13,Gallerani17,SmithBromm2019,Inayoshi19}. 

Growth from stellar-mass BH remnants of Population~III (Pop~III) stars \citep[e.g.][]{Madau01,Abel02,Heger02,Tan04,VolonteriRees2006,McKee08,Yoshida08,Clark11Sci,Greif11,Susa14,Hirano14,Stacy16} to SMBHs is difficult because the gas accretion rate is suppressed by 
radiative and kinetic feedback processes \citep{Whalen+2004,Milosavljevic09,Alvarez09,Tanaka09,Tanaka12,Regan19} 
and growth by mergers is made inefficient by large recoil induced by gravitational wave emission during mergers, which unbinds the merger remnant BHs from the shallow potential wells of their early hosts~\citep{Haiman2004}. These difficulties have motivated several alternative pathways.

One pathway is the direct collapse of supermassive stars (SMSs) \citep[e.g.][]{Omukai01,Oh02,Bromm03,Begelman06,Spaans06,Shang10,Hosokawa12,Agarwal12,Latif13,Sugimura14,Inayoshi14,Ferrara14,Tanaka14,Becerra15,Chon+2016,Hosokawa16,Umeda16,Hirano+2017,Haemmerle18}. If a gas cloud in a massive halo with virial temperature $T\gsim 8000\,$K has no metals or ${\rm H_2}$ molecules, the gas cloud can collapse without fragmentation and grow to become an SMS \citep{Oh02,Volonteri05}. 
However, the background UV radiation flux required to prevent ${\rm H_2}$ molecule formation is 
as high as a few times $10^4$ in units of $J_{21}$ \citep[see, e.g.][and references therein]{WolcottGreen19} 
because of the high density reached via atomic cooling~\citep{Omukai01,Oh02} and self-shielding of ${\rm H_2}$ for realistic UV spectra produced by Population~II stars \citep{WolcottGreen2011b,Sugimura14,Agarwal15,WolcottGreen2017}. The condition of such a strong background radiation is satisfied only in rare cases, in collapsing halos that have bright nearby neighbors \citep{Dijkstra+2008}.  While this is a rare special configuration, it appears feasible for a sufficient number of such pairs of halos to form nearly simultaneously~\citep{Visbal+2014}, while avoiding metal pollution \citep{Dijkstra14}, tidal disruption~\citep{Chon+2016}, and photoevaporation~\citep{Regan+2017}.  For gas in halos located in regions of unusually high baryonic streaming motions~\citep{Hirano+2017}, and/or in halos with unusually rapid merger histories experiencing compressional heating~\citep{Yoshida+2003,Fernandez+2014,Inayoshi+2018}, the UV flux required to avoid ${\rm H_2}$ cooling can be significantly reduced~\citep{Wise+2019}.
  
A second possible pathway is hyper-Eddington accretion onto a stellar-mass BH \citep[][]{Begelman79,Volonteri05,Pacucci15,Inayoshi16,Sakurai16,Pacucci17,Sugimura18,Takeo18,Toyouchi19}. 
Here the problem is that inefficient angular momentum transfer is estimated to reduce the accretion rate (\citealt{Inayoshi18,Sugimura18}, but see \citealt{Alexander13} for a possible solution if the seed BH is surrounded by a massive and dense star cluster).
Then the accretion of an isothermal rotating disk \citep{Oh02} may not be rapid enough to increase the mass of a BH by several orders of magnitude \citep{Sugimura18}. 
Also kinetic feedback may limit the growth rate of BHs \citep{Regan19}. 

A third possibility is runaway mergers of stars and stellar remnants in dense clusters \citep[e.g.][]{PortegiesZwart99,PortegiesZwart02,PortegiesZwart04,Gurkan04,Rasio04,Omukai08,Devecchi09,Vanbeveren09,Glebbeek09,Davies11,Fujii13,Lupi14,Katz15,Tagawa15,Tagawa16,Yajima16,Sakurai17,Sakurai18,Nakauchi18,Boekholt18,Reinoso18,AlisterSeguel19}. 
In high-density stellar systems, 
$\sim 10^{3-4}\,{\rm M_\odot}$ BHs can form in the cluster's center \citep{Omukai08,Devecchi09,Katz15,Sakurai17}. However the seed mass of $\sim 10^3\,{\rm M_\odot}$ may not be massive enough to grow into the SMBHs observed at $z\sim6$ \citep{diMatteo12,Regan19}. 
Thus it is still debated how high-$z$ SMBHs could have formed.

In this study, we focus on environments similar to the direct collapse scenario.  A significant caveat of this scenario is that the
collapsing gas may fragment efficiently, resulting in the formation of a cluster of stars, preventing 
the gas from fueling the formation of a central SMS.  This would then lead to a third pathway, which is expected to produce BH remnants with masses $\approx 10^{3-4}\,{\rm M_\odot}$. 
On the other hand, we propose here that if stars themselves continue to be accreted efficiently, a more massive SMS may form despite the fragmentation of the parent
cloud.  In order for this to occur, incoming stars must collide with
the central star in sufficiently rapid succession such that the
central star never has time to cool and contract and settle on the
main sequence.  This scenario is similar to the runaway mergers
above, but differs in detail. Stars are brought to the central
region of the halo by both gas and stellar dynamical friction.  The
central SMS bloats up to $\gtrsim$astronomical unit size, facilitating the continued
accretion of other stars.  To distinguish this from the usual
``runaway merger'' case, we refer to this variant as ``stellar
bombardment''.  To investigate the feasibility of such a scenario, we
have performed numerical modeling, incorporating star formation,
dynamical friction by gas and stars, gas accretion, stellar
collisions, and gas ejection.

After the submission of our paper, \citet{Chon20} presented results investigating similar scenarios (collapsing gas with a small amount of metal pollution without $\rm H_2$ molecules) using three-dimensional hydrodynamical simulations. They find that for metallicities up to  $10^{-4}\,{\rm Z_\odot}$, a central star can keep growing to $\sim 10^4\,\Msun$ over $\sim 10^4\,\mathrm{yr}$ with high growth rate due to gas accretion and stellar accretion. Their results confirm the expectations in this paper.

\section{Physical picture}

The SMBHs of $\sim 10^9\,\Msun$ at $z\sim6$ are rare objects with an abundance $\sim1\,\mathrm{Gpc^{-3}}$, and thus rare conditions 
may be required to explain their formation 
\citep[e.g.][]{Buchner19}. 
The situation we consider is similar to the usual direct collapse scenario \citep[e.g.][]{Bromm03, Shang10}. 
In this scenario, ${\rm H_2}$ molecules are disrupted in the collapsing cloud by strong background radiation from nearby galaxies, the host halo is massive, and the collapsing cloud is not polluted by metals. 
These conditions keep the cloud at a high temperature, and so enable the cloud to collapse into an SMS without fragmentation. 
Recent studies have suggested that it may be difficult to satisfy these conditions \citep{Latif15}, particularly because a large ${\rm H_2}$-dissociating flux may be required for an extended period, prior to reaching the ``atomic cooling'' threshold~\citep{Regan+2017}.  This may be alleviated only in rare overdense regions, via dynamical heating accompanying unusually rapid merger histories~\citep{Wise+2019}.

Here, instead, we relax the assumption of (the lack of) metal pollution. We consider a massive host halo, a moderate amount of metal pollution of $\sim10^{-5}\,{\rm Z_\odot}$, 
and no ${\rm H_2}$ molecules in a collapsing cloud. 
In such environments, fragmentation only occurs in high-density regions of $\sim 10^{11}\,\mathrm{cm^{-3}}$ due to weak cooling by a small amount of dust grains \citep{Omukai08,Latif16}. After an ultrahigh-density star cluster forms via gas fragmentation, runaway mergers can proceed. In this process, the final mass of the central object is constrained by radiation and supernova (SN) feedback onto a collapsing cloud from newly formed stars, 
since if gas was ejected by feedback, the central object could grow at most to some fraction of the masses of stars (and compact objects) in the cluster. 

The main feedback processes from stars are photoionizing UV radiation and/or SN explosions, which can eject gas from the host halo \citep{Whalen+2004,Kitayama04,Kitayama05}. 
Photoionization feedback from a star influences gas on large scales when the Str{\" o}mgren radius $R_{\mathrm{St},i}=(3{Q}_{\mathrm{ion},i} /4\pi n_\mathrm{gas}^2 \alpha_\mathrm{rec,B})^{1/3}$ exceeds the effective Bondi radius $R_{\mathrm{eff,B},i} \equiv (Gm_i/c_\mathrm{s}^3)(1-L_i/L_{\mathrm{E},i})$.
Here $R_{\mathrm{eff,B},i}$  is the radius within which ambient gas is bound to the star and is modified from the standard Bondi radius to incorporate radiation pressure to ionized gas \citep{McKee08}, $\alpha_\mathrm{rec,B}$ is the case-B recombination coefficient for H (evaluated at $T=10^4$~K), $n_\mathrm{gas}$ is the gas number density, $G$ is the gravitational constant, $c_\mathrm{s}$ is the sound velocity of gas, 
${Q}_{\mathrm{ion},i}$ is the ionizing photon number flux emitted from a star, $m_i$ is the mass of a star, $L_i$ is the luminosity of a star, $L_{\mathrm{E},i}$ is the Eddington luminosity, and subscript $i$ represents the $i^{\rm th}$ star in the cluster.

For main-sequence stars of $\lesssim 30\,{\rm M_\odot}$ in high gas density environments of $\sim 10^{11}\,\mathrm{cm^{-3}}$, the Bondi radius always exceeds the Str{\" o}mgren radius~\citep{Tan04,Hosokawa12,McKee08}. 
Thus, photoionization feedback from the low-mass stars of $\sim 0.1-1 M_\odot$, expected to be born from metal-poor gas \citep{Omukai08,Dopcke13} cannot quench accretion and star formation unless these stars grow to $\sim 30\,\Msun$ by gas accretion or mergers.  Furthermore, photoionization feedback from a massive star becomes efficient only after the star contracts \citep{Hosokawa12}. Stars typically contract on the Kelvin Helmholtz (KH) timescale, which is the timescale for a star to radiate away its gravitational binding energy. On the other hand, \citet{Hosokawa12} and \citet{Haemmerle18} have shown that when the accretion rate onto a protostar exceeds a critical rate of ${\dot m}_\mathrm{cri}\sim (0.006-0.03)\,{\rm M_\odot}/\mathrm{yr}$, the protostar continues expanding, because the heating rate of its envelope due to gas accretion exceeds the radiative cooling rate. 
The production rate of ionizing photons emitted by the soft spectrum of the bloated star is so low that the gas dynamics is not influenced by photoionization feedback \citep{Kitayama04}. 
\citet{Sakurai15} have found that even when there are quiescent phases of accretion onto protostars, they keep expanding if the time-averaged accretion rate within the KH timescale (evaluated at the stellar surface) exceeds ${\dot m}_\mathrm{cri}$. 

The above suggests that if the growth rate of massive stars by mergers with other stars, averaged on the KH timescale, exceeds ${\dot m}_\mathrm{cri}$, massive stars would continue expanding for the same reason.  The growth of massive stars may remain efficient 
in this way, until gas is ejected by an SN explosions of 
one of the massive stars or by accretion feedback from a collapsed massive BH.  Thus, there is a possibility that efficient stellar accretion may help to keep the stellar envelope expanding and so inhibit strong feedback from a contracting massive star, thereby leading to the formation of an SMS.
In the rest of this paper, unless specified otherwise, the expression 
``stellar accretion'' refers to the central protostar colliding and merging with other stars in the core of the halo. 

In this paper, we calculate the evolution of stars that form in high gas density environments as predicted in \citet{Omukai08}. 
\citet{Sakurai17} calculated the evolution of stars formed in a massive halo. 
While they assumed that some fraction of gas is converted to stars at the beginning of the simulation and at the same time gas is ejected, in this paper we consider the evolution of stars 
including the effects of continuous star formation. 
Our pathway is similar to the situation in \citet{Boekholt18}, who calculated collisions of accreting stars. 
In their model, stars are assumed to be kept in the expanded phase due to high gas accretion rates of $\gsim 0.03\,\Msun/\mathrm{yr}$ per star, 
and dynamical interactions are restricted to the two-body relaxation among stars, while hydrodynamical interactions with gas are neglected. 
\citet{Boekholt18} and \citet{Reinoso18} find that the efficiency of collisions increases as the radii of the stars grow. 
On the other hand, we find that even when the gas accretion rate is limited by the Bondi-Hoyle-Lyttleton accretion rate, the central star can keep expanding due to accretion of stars.

We find that the central star can grow efficiently due to the following feedback loop. 
First, stars are captured by the central star by efficient migration due to stellar dynamical friction. 
The radius of the central star then grows, both because of its increase in mass, and because of the heating of its envelope by stellar accretion.
Due to the larger stellar radius, more stars can be captured by the central star. 
Thus stellar accretion is facilitated by both the mass segregation due to dynamical relaxation processes and the growth of the stellar radius due to stellar accretion. 
As mentioned above, we refer to this growth process by stellar accretion as ${\it stellar\, bombardment}$ to distinguish it from the usual runaway collisions. 

In the usual runaway collisions, only a small fraction of stars in the cluster forms a core and the core collapses. 
The core is maintained due to the heating by hard binaries, whose binding energy can be a large fraction of the binding energy of the cluster \citep[e.g.][]{PortegiesZwart99}. 
On the other hand, during stellar bombardment, the binding energy of hard binaries is a tiny fraction of the binding energy of the cluster, since the central star can be expanding during the evolution, so stars can accrete onto the central star almost without being heated by hard binaries. Thus both the dynamical evolution of surrounding stars and the final outcome (i.e. the mass of the central BH remnant) are qualitatively different between the runaway collision and the stellar bombardment.

\begin{figure}
\begin{center}
\includegraphics[width=85mm]{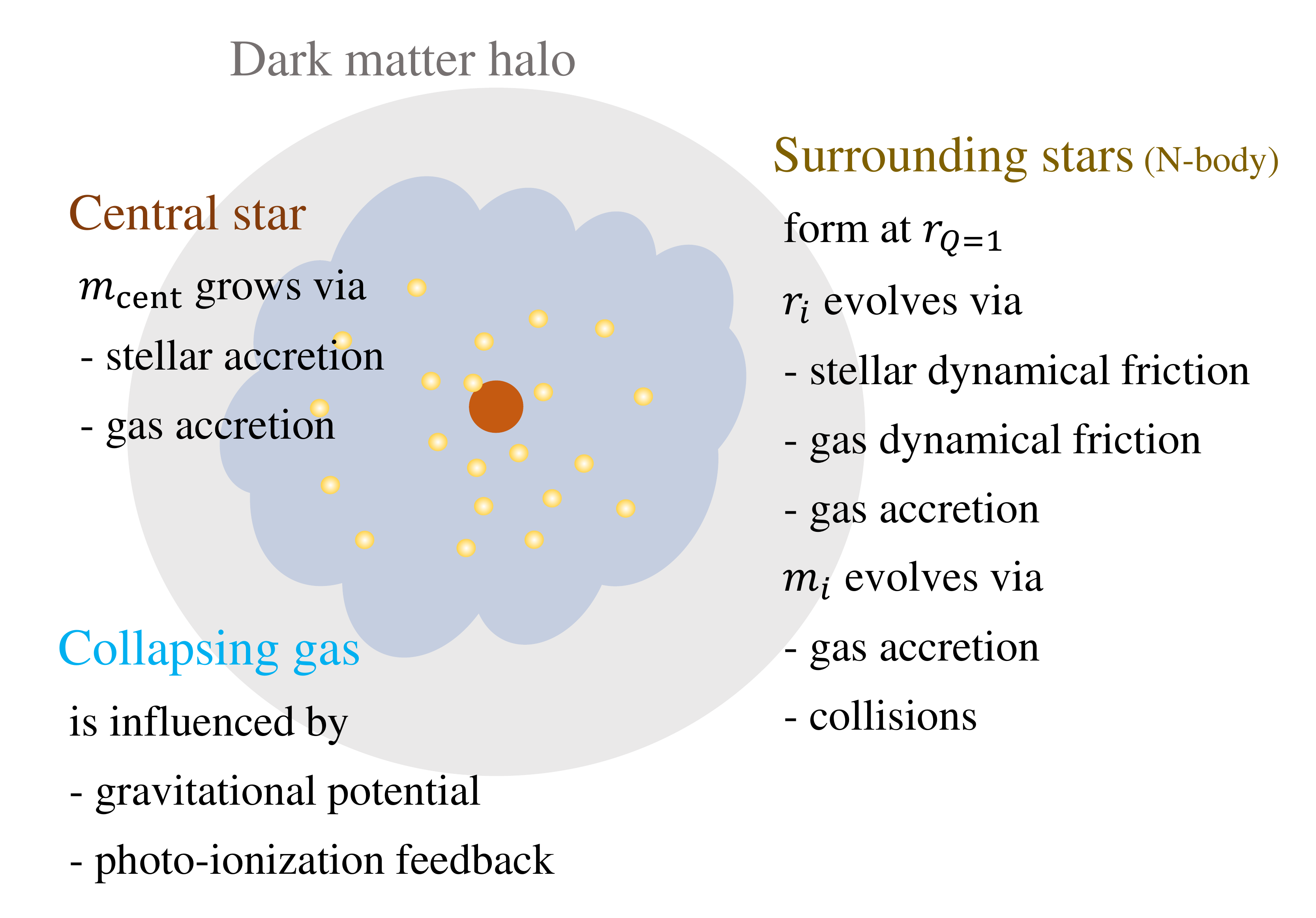}
\vspace{\baselineskip}
\caption{
Schematic diagram illustrating the system we study and the mechanisms affecting each of its components. Surrounding stars are characterized by their radial position ($r_i$) from the central star and their mass ($m_i$).  
These variables are updated via semianalytic prescriptions in each ``$N$-body'' time step, due to multiple processes as listed in the diagram. 
}
\label{fig:method_diagram}
\end{center}
\end{figure}

\section{Method}

To investigate how stars form, migrate inward, and crash into the central star, and how they are affected by feedback, we use a semianalytical model incorporating the effects of 
star formation, protostellar evolution, gas accretion, dynamical friction by stars and gas, collisions, and gas ejection (Fig.~\ref{fig:method_diagram}). In this section, we provide an overview of our simulations.

\subsection{Setup and initial conditions}

We consider the following components: a central star, surrounding stars, a gas cloud, and a dark matter (DM) halo. 
The label ``surrounding stars'' refers to all stars other than the central star.  We reserve the term ``stars'' throughout the
paper to include both surrounding stars and the central star. 
We follow the evolution of the entire system for 3~Myr, which is roughly the time when either the first surrounding star may be expected to explode or the central star collapses to a BH.

In our semianalytical model, $N$-body particles represent surrounding stars. Surrounding stars form, migrate, and accrete onto the central star, while the central star is pinned to the center of the system 
neglecting both gas driven migration and wandering due to dynamical interactions with other stars.\footnote{Assuming that the central star is pinned to the center is not a major simplification. While in reality it may wander away from the center due to dynamical two-body interactions with the surrounding stars, as long as it has a mass lower than or comparable to that other stars in the cluster, generally the most massive objects in the cluster sink to the central region due to the {\it Spitzer} instability and become prone to stellar collisions.
In this case, we assume that the most massive star becomes the central star. 
} 
We investigate several values for the maximum initial mass of surrounding stars ($m_\mathrm{0,max}$), 
and also set the initial mass of the central star to be $m_\mathrm{cent}=m_\mathrm{0,max}$. 
As a fiducial model, we set $m_\mathrm{0,max}=1\,{\rm M_\odot}$, 
which is roughly the Jeans mass at which fragmentation occurs at the density $10^{10}\,\mathrm{cm}^{-3}$ \citep{Omukai08}. 
We assume that there are no surrounding stars initially.

When stars are expanding, we set their radius to 
\begin{align}
\label{eq:r_expand}
R_i=2.6\times 10^3\,R_\odot (m_i/100\,\Msun)^{0.5}, 
\end{align}
following \citet{Hosokawa12}, 
while after stars are contracted (the condition of these stars depends on their accretion rate and the KH timescale as described in \S~\ref{sec:stellar_evolution} below), 
their radius is assumed to be
\begin{align}
\label{eq:r_contract}
R_i= R_{\mathrm{ZAMS},i} =4.6\,R_\odot (m_i/100\,\Msun)^{0.58} 
\end{align}
\citep{Hirano17b}. 
Throughout this paper, we refer to a star in the expanding (pre-main-sequence) and the contracting (main-sequence) phases as an ``expanding star'' and a ``contracted star'', respectively.

\subsubsection{Density profile}
\label{sec:gas_model}

We set the number density profile for gas $n_\mathrm{gas} (r)$ as 
\begin{equation}
\label{eq:ngas}
n_\mathrm{gas} (r)=
\left\{
\begin{array}{cl}
     n_\mathrm{c} &\mathrm{if}~r \leq r_\mathrm{c}  \\
     n_\mathrm{c} (r/r_\mathrm{c})^{-2}
     &\mathrm{if}~ r_\mathrm{c}\leq r \leq r_{\rm vir}\\
     0 &\mathrm{if}~r> r_{\rm vir}
\end{array}
\right.
\end{equation}
where $r_\mathrm{c}$ is the core radius of the collapsing gas and $n_\mathrm{c}$ is the core gas density. 
Outside the core radius, gas is assumed to be collapsing under its self-gravity, while in the dense core, gas is assumed to cool efficiently, fragment, and form stars. 
As a fiducial model, we set $n_\mathrm{c}=10^{11} \,\mathrm{cm}^{-3}$, and the temperature of inflowing gas to $T=10^4\,\mathrm{K}$ \citep[e.g.][]{Oh02,Omukai08,Shang10}. Assuming an isothermal equation of state, the sonic velocity of inflowing gas is $c_\mathrm{s}= ( kT / \mu )^{1/2} \simeq10\,\mathrm{km/s} (T/10^4\,\mathrm{K})^{1/2}$, where $k$ is the Boltzmann constant and $\mu=1.22$ is the mean mass per particle, and the accretion rate from large scales is set to 
\begin{equation}\label{eq:Mdotin}
{\dot M}_\mathrm{in} = \frac{c_\mathrm{s}^3}{G} \simeq 0.22\,\frac{\Msun}{\mathrm{yr}} \left(\frac{T}{10^4\,\mathrm{K}}\right)^{3/2} \end{equation}
\citep[e.g.][]{Stahler80,Begelman06}. 
The core density $n_\mathrm{c}= 10^{11} \,\mathrm{cm}^{-3}$ roughly matches the density at which gas with a metallicity of $\sim10^{-5}\,Z_\odot$ and a suppressed ${\rm H}_2$ fraction (by strong background radiation) begins to fragment due to the decrease of its temperature~\citep{Omukai08}. In the fiducial model, the initial value of the core radius for the collapsing gas is chosen to be $r_\mathrm{c,ini} \simeq 4 \times 10^{-4}$ pc by matching the cosmological baryon-to-DM mass ratio inside the virial radius $r_\mathrm{vir}=(3M_\mathrm{halo}/800 \pi \rho_\mathrm{cri})^{1/3}$, where $M_\mathrm{halo}$ is the halo mass within the virial radius, 
$\rho_\mathrm{cri}=3H^2/8\pi G$ is the cosmological critical density, $H\simeq H_0[\Omega_\mathrm{m0} (1+z)^3+\Omega _{\Lambda0}]^{1/2}$ is the Hubble parameter, $H_0 \simeq 70\,\mathrm{km/s/Mpc}$ is the Hubble constant,  $\Omega_\mathrm{m0}=0.24$ is the matter density today, and $\Omega _{\Lambda0}=0.76$ is the cosmological constant today \citep{Plank16}. 
We assume that the halo mass within the virial radius $M_\mathrm{halo}$ is $10^{7}\,\Msun$.
The radius at which $n_\mathrm{gas}(r)= 10^{11} \,\mathrm{cm}^{-3}$,  measured in high-resolution cosmological hydrodynamical simulations of metal- and ${\rm H_2}$-free gas in atomic-cooling halos \citep[e.g.][]{Regan14} is also found to be $\sim 10^{-4}-10^{-3}$ pc. 
We also checked that our results are not significantly influenced by changing the value of $r_\mathrm{c,ini}$ to $10^{-3}$ pc, which is because $r_\mathrm{c}$ quickly evolves 
due to our assumption of setting $r_c$ to the place where the gas becomes unstable to fragmentation (see \S \ref{sec:star_formation}). 
Thus, we assume that the core gas density is fixed while $r_\mathrm{c}$ evolves with time (\S \ref{sec:star_formation}). 
In our model, the final results depend on the position of star formation at $r_\mathrm{c}$, while they are less affected by other effects related to the gas density distribution.\footnote{ The main effects of the background gas distribution in the simulation are (i) to generate a potential that influences the stellar collision probability (\S~\ref{sec:mergers}), and (ii) to drive stellar bombardment through gas dynamical friction (\S~\ref{sec:gdf}). The growth rate of the central star is not directly influenced by $n_{\rm gas}$ since we limit the stellar mass accretion rate and star formation by the inflowing gas supply rate from the outer regions $\dot{M}_\mathrm{in}$ (\S~\ref{sec:accretion} and ~\ref{sec:star_formation}).} Therefore we expect that the core gas density fixed in time is not a critical assumption.

When we calculate the acceleration due to gas dynamical friction and accretion (\S \ref{sec:gdf} and \ref{sec:accretion}), 
we assume that the gas mean velocity is zero for simplicity. 
This assumption gives an optimal rate for migration toward the center for surrounding stars due to gas dynamical friction and accretion. Nevertheless, the migration by gas dynamical friction and accretion is found to give small contributions to the final SMS mass (\S \ref{sec:central_evolution}).

\subsubsection{Gravitational potential}

We adopt a four-component gravitational potential,
\begin{eqnarray}
\Phi (r) =
\Phi_\mathrm{DM}(r)+\Phi_\mathrm{star}(r)+\Phi_\mathrm{gas}(r)+\Phi_\mathrm{cent}(r), 
\end{eqnarray}
where $\Phi_\mathrm{DM}(r)$, $\Phi_\mathrm{star}(r)$, $\Phi_\mathrm{gas}(r)$, and $\Phi_\mathrm{cent}(r)$ are, respectively, the gravitational potential at the position $r$ of the dark matter halo, surrounding stars, collapsing gas, and the central star. 

We set $\Phi_\mathrm{DM}(r)$ by the Navarro-Frenk-White (NFW) profile as 
\begin{eqnarray}
\Phi_\mathrm{DM} (r)=\frac{-4\pi G\rho_\mathrm{h} r_\mathrm{h}^3}{r} \ln\left(\frac{r_\mathrm{h}+r}{r_\mathrm{h}}\right),
\end{eqnarray}
where $r_\mathrm{h} = r_\mathrm{vir}/C$ is the scale radius of the halo, 
$\rho_\mathrm{h} = 200 \rho_\mathrm{cri} C^3/(3\ln(1+C)-C/(1+C))$ is the density parameter of the NFW profile, and $C$ is the concentration parameter
\citep{Navarro97}. 
We assume that $C=9$. 
We assume a redshift $z=15$,
since atomic-cooling halos (whose masses are
$\approx 10^{7}\,\Msun$ at this redshift) start to appear from around this epoch \citep[e.g.][]{Tanaka09}. 
For reference, we also note that 
in the context of trying to grow to an SMBH of $\sim 10^9\,\Msun$ at $z\sim 6$ via gas accretion, 
the seed BH mass is required to be $\sim10^5\,\Msun$ at $z\sim 10$  \citep{diMatteo12}.

The gravitational potential of the gas is derived from equation (\ref{eq:ngas}) using Eq.~(2.28) in \citet{Binney08} as 
\begin{equation}
\Phi_\mathrm{gas} (r)=
4\pi G n_\mathrm{c} \mu m_\mathrm{H}\times\left\{
\begin{array}{l}
\dfrac{r^2}{6}-\dfrac{r_\mathrm{c}^2}{2} - r_{\mathrm{c}}^2 \ln  \dfrac{r_\mathrm{vir}}{r_{\mathrm{c}}}\\
\qquad\qquad \mathrm{for}~r<r_\mathrm{c}\,,\\
\dfrac{2 r_{\mathrm{c}}^3}{3 r} - r_{\mathrm{c}}^2 -  r_{\mathrm{c}}^2 \ln \dfrac{r_\mathrm{vir}}{r}  \\
\qquad\qquad\mathrm{for}~r>r_{\mathrm{c}}\,, 
\end{array}
\right.
\end{equation}
where $m_\mathrm{H}$ is the hydrogen mass. 

Similarly, for the gravitational potential of the surrounding stars: 
\begin{eqnarray}
\Phi_\mathrm{star} (r)=-\frac{GM_*(<r)}{r}-\int^{R_\mathrm{max}}_r 4\pi r G\rho_* (r) dr ,
\end{eqnarray}
where $\rho_* (r)$ is the stellar density at $r$, and $M_*(<r)$ is the stellar mass within $r$. 
We integrate and derive the gravitational potential at the center of the spherical radial cell $l$ in each time step, 
using a linear interpolation of 
$\Phi_\mathrm{star}(r)$. 
We use 60 radial cells, covering the range from $10^{-8}$ pc to $R_\mathrm{max}=10$ pc, spaced uniformly in log~$r$.

The gravitational potential by the central star is 
$\Phi_\mathrm{cent}(r)=-Gm_\mathrm{cent}/r$. 

Note that the profiles of $\Phi_\mathrm{DM}(r)$, $\Phi_\mathrm{gas}(r)$, and $\Phi_\mathrm{cent}(r)$ are assumed to be algebraically fixed as given above, while $r_\mathrm{c}$ and the normalizations of $\Phi_\mathrm{gas}(r)$ and $\Phi_\mathrm{cent}(r)$ are followed in time. 
$\Phi_\mathrm{star}(r)$ is assumed to evolve, and its full radial profile is followed during our calculations.

\subsection{Star formation}
\label{sec:star_formation}

We assume that the any gas flowing in from large scales that is not accreted onto stars is converted into new surrounding stars, and thus the star formation rate is 
\begin{equation}\label{eq:mdot_SF}
{\dot m_\mathrm{SF}}={\dot M}_\mathrm{in}-\sum_i{\dot m}_{i}    
\end{equation}
where the sum goes over all stars, including the central star and surrounding stars. 
This assumption is not obvious, since the inflowing gas may accumulate in the central region and increase the core density and core radius. Since it is difficult to follow the time evolution of the core density due to this gas accumulation, we assume a constant core density and a constant inflow rate as input parameters.
Due to the temperature dependence of the inflow rate given by Eq.~\eqref{eq:Mdotin}, cases with low star formation rates are investigated in models with low $T$ below. 

Following simulations of Population~II star formation by \citet{Dopcke13}, we set the initial mass function to 
\begin{equation}\label{eq:massfunction}
    \frac{dN}{dm} \propto m^{-\beta} ~~\mathrm{where}~~ 0.08 M_\odot \leq m \leq m_\mathrm{0,max}
\end{equation}
with a flat logarithmic distribution, $\beta=0$, as a fiducial model. 
In practice, in each time step we form new stars randomly from this mass function in succession as long as their total mass is less than 
$m_\mathrm{SF}={\dot m_\mathrm{SF}}\Delta t$. Let us label with $n$ the corresponding largest number of new stars where this criterion holds. Finally we form the last star with probability
$m_{\mathrm{new},n+1}/(m_\mathrm{SF}-\Sigma_{i=1}^n m_{\mathrm{new},i})$, where the mass of the ${n+1}^{\rm th}$ newly formed star $m_{\mathrm{new},n+1}$ is drawn randomly from the mass function.

We assign cells to the newly formed stars based on the following arguments. 
Although we assume that the collapsing gas cloud is overall spherically symmetric, gas disks may form around the central star or around surrounding stars forming and growing by accreting gas with nonzero angular momentum. 
Since a gas disk is stabilized by rapid rotation in a steep gravitational potential, surrounding stars form outside the radius where the Toomre parameter 
\begin{equation}
\label{eq:q_value}
Q=\frac{c_\mathrm{s} \Omega}{\pi G \Sigma_\mathrm{gas}}=\frac{ \Omega_\mathrm{Kep}\Omega}{2\pi G n_\mathrm{gas} \mu m_\mathrm{H} }
\end{equation}
becomes 1, 
where $\Omega$ is the orbital frequency of the gas disk, and $\Omega_\mathrm{Kep}(r)=(GM(r)/r^3)^{1/2}$ is the Keplerian orbital frequency, where $M(r)$ is the enclosed mass,
and $\Sigma_\mathrm{gas}$ is the surface density of the gaseous disk. 
Since gas disks are partially supported also via turbulent and thermal pressure, $\Omega = \epsilon_\mathrm{Kep} \Omega_\mathrm{Kep}$ where $\epsilon_\mathrm{Kep}$ describes the deviation from a fully rotationally supported disk. Referring to the results of simulations for primordial disks \citep[e.g.][]{Greif12,Latif13,Hirano14}, we adopt $\epsilon_\mathrm{Kep}=0.5$. 
In the case of $r_{Q=1}>r_\mathrm{c,ini}$, unless the total accretion rate is as high as $\sim {\dot M}_\mathrm{in}$, gas accumulates within the central region. Unlike our simplified assumption of a homogeneous $n_{\mathrm{gas}}$, a more realistic density distribution for Keplerian rotating gas is $\rho\propto r^{-1/2} - r^{-3/2}$ \citep[e.g.][]{Inayoshi18}. However, even for this density profile within the core, gas is most unstable in the outer region where $n_\mathrm{gas}\geq n_\mathrm{c}$. 
Thus we assume that surrounding stars form at $r_{Q=1}$ when $r_{Q=1}>r_\mathrm{c,ini}$, 
while surrounding stars form uniformly from $r_{Q=1}$ to $r_\mathrm{c,ini}$ when $r_{Q=1}<r_\mathrm{c,ini}$. 
Accordingly the core radius is set to $r_\mathrm{c}=r_{Q=1}$ in each time step using equation~\eqref{eq:q_value}. 
From equation (\ref{eq:q_value}), $r_\mathrm{c}=r_{Q=1}$ is satisfied at 
\begin{equation}
\label{eq:rc_m}
r_\mathrm{c}=\left[ \frac{\epsilon_\mathrm{Kep}}{\left(\frac{3}{2} - \epsilon_\mathrm{Kep}\right)} \frac{[m_\mathrm{cent} + M_\mathrm{stars}(r_\mathrm{c}) + M_\mathrm{DM}(r_\mathrm{c})]}{\frac{4}{3}\pi n_\mathrm{c} \mu m_\mathrm{H}}\right]^{1/3}.
\end{equation}
Let us introduce $M_\mathrm{DM}(r)$ and $M_\mathrm{stars}(r)$ to label the enclosed mass of the dark matter and surrounding stars within $r$, respectively. The dark matter mass is typically subdominant in this expression. In the early phases, the total stellar mass of surrounding stars and the central star is limited by the inflow rate from large scales and  $m_\mathrm{cent}+M_\mathrm{stars}(r)\sim {\dot M}_\mathrm{in}t$, 
which implies that
\begin{equation}
\label{eq:rc_t}
r_\mathrm{c} \approx \frac{\epsilon_\mathrm{Kep}^{1/3}}{\left(\frac{3}{2}-\epsilon_\mathrm{Kep}\right)^{1/3}} \frac{\dot{M}_\mathrm{in}^{1/3}t^{1/3}}{\left[\frac{4}{3}\pi n_\mathrm{c} \mu m_\mathrm{H}\right]^{1/3}}.
\end{equation}
We ignore the dynamical effect on the surrounding stars due to the deepening of the gas gravitational potential if $r_\mathrm{c}$ moves outwards, which tightens the orbit of surrounding stars outside of $r_\mathrm{c}$.

In order to form stars, cooling from dust grains needs to be stronger than heating, since gas fragmentation is caused by the decrease of the gas temperature due to cooling by dust grains \citep{Omukai08}. 
In our model, gas is heated by gas dynamical friction, 
with the total heating rate given by
\begin{equation}
\label{eq:heating_gdf}
\Gamma_\mathrm{GDF}=
\frac{\mathrm{d}E_\mathrm{GDF}}{\mathrm{d}t}
=\sum_i m_i v_i a_{\mathrm{GDF},i}.
\end{equation}
We assume that even when gas dynamical friction does not reduce the velocity of surrounding stars at 
$\Sigma_{i \in l}4 \pi r_{\mathrm{Bondi},i}^3/3 > V_l$ (\S \ref{sec:gdf}), 
gas is heated by gas dynamical friction, 
and we substitute $a_{\mathrm{GDF},i}$ calculated by equation (\ref{eq:gdf}) into equation (\ref{eq:heating_gdf}). 
Thus equation (\ref{eq:heating_gdf}) represents the upper limit for the heating rate due to gas dynamical friction.

Referring to \citet{Omukai08}, the specific cooling rate from dust grains is 
\begin{align}
\label{eq:cooling_dust}
\lambda_\mathrm{dust}\sim 
\left\{
\begin{array}{l}
100\,\dfrac{\mathrm{erg}}{\mathrm{s \,g}} \dfrac{n_\mathrm{gas}}{10^{10}\,\mathrm{cm^{-3}}}  
\\
\qquad\mathrm{for}~ 10^6\,\mathrm{cm^{-3}}<n_\mathrm{gas}<10^{10}\,\mathrm{cm^{-3}}, \\[1.5ex]
100\,\dfrac{\mathrm{erg}}{\mathrm{s\, g}}
\left(\dfrac{n_\mathrm{gas}}{10^{10}\,\mathrm{cm^{-3}}}\right)^{0.2} \\
\qquad\mathrm{for}~ 10^{10}\,\mathrm{cm^{-3}}<n_\mathrm{gas}<10^{12}\,\mathrm{cm^{-3}}.
\end{array}
\right.
\end{align}
Since the core region is optically thin to dust emission \citep{Omukai08}, the net cooling rate by dust grains within the core region is $\Lambda_\mathrm{dust}=\lambda_\mathrm{dust}M_\mathrm{c}$, where $M_\mathrm{c}=\frac{4 \pi}{3}n_\mathrm{c} \mu m_\mathrm{H} r_\mathrm{c}^3 $ is the gas mass within the core radius. Whenever the heating rate exceeds the cooling rate, 
$\Gamma_\mathrm{GDF}>\Lambda_\mathrm{dust}$, 
star formation is assumed to be quenched. 
When the cooling dominates the heating, the gas temperature is expected to decrease and gas fragments as found in \citet{Omukai08}. 

\subsection{Expanding and contracting stars}
\label{sec:stellar_evolution}

For each star, we specify whether they are in the expanding (pre-main-sequence) or contracting (main-sequence) phase in each simulation time step as follows.
In isolation, a protostar contracts on the KH timescale  
$t_{{\mathrm{KH},i}}=Gm_i^2/(R_iL_i)$. 
On the other hand, \citet{Hosokawa12} have shown that if the mass accretion rate (see \S~\ref{sec:accretion}) exceeds the critical rate, 
$\dot{m}_i > {\dot m}_\mathrm{cri}\sim0.006-0.03~\Msun/\mathrm{yr}$, stars can keep expanding (see also \citealt{Haemmerle18} for similar results). Furthermore, \citet{Sakurai15} have shown that if the mass accretion rate time-averaged over a KH timescale evaluated on the stellar surface, $t_{\mathrm{surf,KH},i}\sim10~t_{{\mathrm{KH},i}}\sim 10~Gm_i^2/R_iL_i$, exceeds the critical rate $\langle \dot{m}_i\rangle > {\dot m}_\mathrm{cri}$, the star can keep expanding. 
Following these results, we assume that if the 
accretion rate $\dot{m}_i$ smoothed on the surface KH timescale 
\footnote{
In practice we define $\langle \dot{m}_i\rangle_{t}$, the smoothed accretion rate of star $i$ at time $t$, recursively using the instantaneous accretion rate $\dot{m}_{i,t}$ as 
\begin{equation}
\label{eq:time_smoothing}
\langle \dot{m}_i\rangle_{t+\Delta t} = \langle \dot{m}_i\rangle_{t} 
\left( 1-\frac{\Delta t}{10t_{\mathrm{KH},i}}\right) + \dot{m}_{i,t} \frac{\Delta t}{10t_{\mathrm{KH},i}}\,. 
\end{equation}
} 
exceeds ${\dot m}_\mathrm{cri} = 0.01~\Msun/\mathrm{yr}$, or the time from its formation is shorter than 
the KH timescale for zero-age main-sequence (ZAMS) stars 
($t_{{\mathrm{KH,ZAMS},i}}=Gm_i^2/(R_{\mathrm{ZAMS},i} L_i)$),
star $i$ keeps expanding, and otherwise it contracts. 
We calculate the surface KH timescale $t_{\mathrm{surf,KH},i}$ using the stellar luminosity $L_i$ given by equations (3), (4), and (5) in \citet{Hosokawa12} for stars with masses of $m_i<6\,\Msun$, $6\,\Msun \leq m_i < 50\,\Msun$, and $m_i \geq 50\,\Msun$, respectively. 
Note that the critical condition (${\dot m}_\mathrm{cri}$) could be lower for accretion of stars than that for accretion of gas. This is because high-velocity accretion of highly eccentric stars and/or the internal energy of accreted stars may heat the envelope of the central star more, per unit infalling mass, compared to accreting cold gas at the same rate.

\subsection{Radial motion}\label{sec:radial}

A particle $i$ (i.e. one of the surrounding stars) 
is described by its mass $m_i$ and its radial distance from the central star $r_i$.  
For simplicity, particles are assumed to follow circular orbits, but they are allowed to migrate radially.
After each time step $\Delta t$, 
we update the position of particle $i$ to $r_i+\Delta r_i$, satisfying
\begin{equation}
\label{eq:migration_rate}
E(r_i+\Delta r_i)=E(r_i)+\Delta k_i 
\end{equation}
where $E(r_i)=\Phi(r_i)+k(r_i)$ is the  total specific energy, 
$k(r_{i})=\frac{1}{2}v_i^2$ is the specific kinetic energy, 
$\Delta k_i$ is the change in the specific kinetic energy within $\Delta t$, and
$v_i$ is the orbital velocity of the $i$th particle. The change in the specific kinetic energy is given as 
\begin{equation}
\Delta k_i=v_i a_i\Delta t
\end{equation}
where $a_i$ is the acceleration of the $i$th particle. We assume $v_i = v_\mathrm{Kep}(r_{i})$, where $v_\mathrm{Kep}(r)$ is the Keplerian orbital velocity at $r$. The acceleration is given as  
\begin{equation}
a_i=a_{\mathrm{SDF},i}+a_{\mathrm{GDF},i}+a_{\mathrm{acc},i}
\end{equation}
where $a_{\mathrm{SDF},i}$,  $a_{\mathrm{GDF},i}$, and $a_{\mathrm{acc},i}$ are the acceleration of the $i$th particle due to stellar dynamical friction, gas dynamical friction, and accretion, respectively (see \S \ref{sec:sdf}, \ref{sec:gdf}, and \ref{sec:accretion} below). 

For simplicity, in the calculation of the migration rate in equation (\ref{eq:migration_rate}), we assume that the eccentricity of a surrounding star does not evolve with time, and remains zero. 
We consider the effects of nonzero eccentricities in 
\S \ref{sec:parameter_depend}.

To follow the migration, we use a shared time step of
\begin{align}
\label{eq:timestep}
\Delta t =\eta \,\mathrm{min}_i \left[ \frac{v_i}{a_{\mathrm{SDF},i}},\frac{v_i}{a_{\mathrm{GDF},i}},
\frac{v_i}{a_{\mathrm{acc},i}},
\frac{1}{\sqrt{G n_\mathrm{c} \mu m_\mathrm{H}}} \right]
\end{align}
where the constant $\eta$ is a time step parameter. On the right-hand side, the four terms are the timescales for stellar dynamical friction, gas dynamical friction, and accretion torque, and the dynamical time within the core radius, respectively. 

We set the time step parameter to be $\eta=0.1$. 
To validate this choice, we compared the final mass of the central star in one of the models (``Model 2'' below) with $\eta=$0.4, 0.2, and 0.1. The mass was found to be 
$4.1\times 10^3\,\Msun$, $3.8 \times 10^3\,\Msun$, and 
$3.7 \times 10^3\,\Msun$, respectively. 
The final mass changes only by $<2\%$ between the last two cases ($\eta=0.2$ and $\eta=0.1$), giving us confidence that our results have nearly converged at $\eta = 0.1$.

\subsubsection{Stellar dynamical friction}

\label{sec:sdf}
Stellar dynamical friction is modeled using the analytic formula \citep{Binney08} of
\begin{align}
\label{eq:sdf}
{a}_{\mathrm{SDF},i}=&-\frac{4\pi G^2 m_i \rho_* \ln\Lambda}{v_i^2}
\left[\mathrm{erf}\left(\frac{v_i}{\sqrt{2}\sigma_*}\right)-
\right.\nonumber\\&\left.
\frac{\sqrt{2}v_i}{\sqrt{\pi}\sigma_* }e^{-v_i^2/2\sigma_*^2}\right], 
\end{align}
where $\sigma_*$ is the velocity dispersion of background stars, $\rho_*$ is the stellar density, $\ln\Lambda=\ln(b_\mathrm{max}/b_\mathrm{min})$ is the Coulomb logarithm, 
and $b_\mathrm{max}$ and $b_\mathrm{min}$ are the maximum and minimum impact parameters for weak stellar encounters. 
We set $b_\mathrm{min}=Gm_i/v_i^2$, $b_\mathrm{max}=0.1$ pc. 
Equation (\ref{eq:sdf}) assumes an isotropic and Maxwellian velocity distribution for background stars. 
We set $\sigma_*=v_\mathrm{Kep}(r_i)/\sqrt{3}$, which sets the value in the square parenthesis in equation (\ref{eq:sdf}) to 0.86 since we assume $v_i=v_\mathrm{Kep}(r_i)$.

To obtain the background density $\rho_*$ at each time step, we compute an average stellar density in each radial cell $l\in[1,60]$, 
obtained from the total number of surrounding stars found in each cell assuming spherical symmetry. 
To check the effect of the number of cells $N_\mathrm{cell}$, we compared the results for Model~2 for $N_\mathrm{cell}=40$, 60, and 80. 
The final mass of the central star in these three cases was found to be $4.1 \times 10^3$, $3.7 \times 10^3$, and $3.6 \times 10^3$, respectively. 
The small difference ($< 3\%$) between the latter two cases gives us confidence that our results nearly converge for $N_\mathrm{cell}=60$.

When $m_i$ is larger than the average mass ($\overline{m}_l$) in some cell $l$ hosting the $i^{\rm th}$ surrounding star, the $i^{\rm th}$ surrounding star migrates inward by the acceleration in equation (\ref{eq:sdf}). On the other hand, when $m_i<\overline{m}_l$, the $i^{\rm th}$ surrounding star gains kinetic energy from the encounter and migrates outward,  which is not accounted for by equation (\ref{eq:sdf}). Due to energy conservation, the total kinetic energy change for all surrounding stars in each cell by stellar dynamical friction is zero.  
To reduce computational time, we assign equal momentum change ($\Delta p_l$) to every below-average surrounding star in each cell.  
Here $\Delta p_l$ is determined from energy conservation by solving  the following equation in each cell
\begin{align}
&\sum_{m_i<\overline{m}_l} \frac{1}{2}m_i 
\left[
  \left( 
     v_i + \frac{\Delta p_{l}}{m_i}
  \right)^2 
  - v_i^2
\right]
\nonumber\\
&\qquad=-\sum_{ m_i>\overline{m}_l} \frac{1}{2}m_i\left[\left(v_i - |a_{\mathrm{SDF},i}| \Delta t \right)^2  -v_i^2\right]\,.
\end{align}
Since $v_i \gg |a_{\mathrm{SDF},i}|\Delta t$ (equation \ref{eq:timestep}) and $\Delta p_l \ll m_i v_i$, we approximate this equation as
\begin{align}
\sum_{m_i<\overline{m}_l} v_i {\Delta p_{l}} =\sum_{ m_i>\overline{m}_l} m_i v_i |a_{\mathrm{SDF},i}| \Delta t \,.
\end{align}
We assign the new radial location to the surrounding stars to match the updated velocity to the circular velocity at that radius (\S~\ref{sec:radial}). 
This procedure ensures that the cluster is in local virial equilibrium everywhere and accounts for two-body relaxation for the stellar cluster in an approximate way.

We assume that stellar dynamical friction operates when the number of surrounding stars within a cell is more than one. 
In the fiducial model, we verify that the number of surrounding stars within $10\,R_\mathrm{cent}$ is more than a hundred at $t=10^4$ yr. Hence the number of surrounding stars is mostly large enough to validate equation (\ref{eq:sdf}) in our models.

\subsubsection{Gas dynamical friction}
\label{sec:gdf}

When a particle has a nonzero velocity relative to the background gas, it suffers additional dynamical friction from the gas component.
Due to this mechanism, surrounding stars may migrate toward the center. 
We use the gas dynamical friction formulation derived by \citet{Ostriker99} as 
\begin{align}
\label{eq:gdf}
a_{\mathrm{GDF},i} &= -\frac{4\pi G^2 m_i n_\mathrm{gas} \mu m_\mathrm{H}}{v_{\mathrm{rel,}i}^2}f(v_{\mathrm{rel,}i}/c_s),\nonumber\\
f(x) &= 
\left\{
\begin{array}{ll}
\frac{1}{2} \mathrm{ln}\left(\frac{1+x}{1-x}\right)-x & \mathrm{for}~~0 < x < 1\,,\\
\frac{1}{2}\,\ln\left(x^2 -1 \right)+\ln\Lambda' & \mathrm{for}~~x>1\,,~~
\end{array}
\right.
\end{align}
where $\ln\Lambda'$ is a Coulomb logarithm for the gas distribution, and $v_{\mathrm{rel,}i}$ is the relative velocity between the $i^{\rm th}$ surrounding star and the background gas. Referring to the result of numerical simulations by \citet{Chapon13}, we adopt $\ln\Lambda'=3.1$. We set $v_{\mathrm{rel,}i}=v_i$ assuming a static background gas distribution.

In the usual formulation of dynamical friction, 
a body is assumed to be moving on a straight line (but see \citealt{KimKim2007,Chapon13}), relative to an unperturbed background. 
Since each star disturbs the gas inside its Bondi-Hoyle-Lyttleton sphere, the formulation of gas dynamical friction is not valid within another star's Bondi-Hoyle-Lyttleton sphere. 
We assume that when the sum of the volumes of the Bondi-Hoyle-Lyttleton spheres of surrounding stars within the spherical cell $l$ ($\Sigma_{i \in l}4 \pi R_{\mathrm{BHL},i}^3/3$) exceeds the volume of the cell $V_l$, gas dynamical friction does not operate in that cell. We likewise neglect gas dynamical friction inside the Bondi radius of the central star.

\subsubsection{Accretion torque}

Due to gas accretion (\S \ref{sec:accretion}), accreted objects receive momentum to satisfy momentum conservation. In this study, we set the acceleration due to gas accretion as 
\begin{equation}
a_{\mathrm{acc},i}=-\frac{\dot{m}_i v_i}{m_i}. 
\end{equation}
For simplicity we assume that gas is static, and the relative velocity between surrounding stars and gas is always given by the velocity of surrounding stars, i.e. the angular momentum is always reduced, which leads to radially inward migration.
Since the collapsing gas may instead have angular momentum in the same sense as the surrounding stars, 
this prescription gives an upper limit for the deceleration and the resulting radial migration rate for surrounding stars. In \S \ref{sec:central_evolution}, we find that the deceleration by gas accretion has a minor effect on the migration of surrounding stars, even at this upper limit.

\subsection{Stellar collisions among surrounding stars}
\label{sec:mergers}

We also consider collisions among surrounding stars. Assuming that the surrounding stars' motion is isotropic, the number, the number density, and the velocity dispersion in cell $l$ are $N_{l}$, $n_{l}$, and $\sigma_{*,l}$, respectively, the expected rate of collisions within the time step $\Delta t$ in a cell $l$ is given by (Eq.~(7.194) in \citealt{Binney08})
\begin{align}
\label{eq:n_coll}
    N_{{\rm coll},l} &= 
    \frac{1}{2} N_{l}\frac{\Delta t}{t_{\mathrm{coll},l}}\nonumber\\
    &=\frac{1}{2} N_{l} n_{*,l} \sigma_{*,l} \left( R_{\rm coll}^2 + \frac{G \overline{m}_l}{\sigma_{*,l}^2}R_{\rm coll} \right)
    \Delta t ,
\end{align}
where 
$R_{\rm coll}$ is the pericenter distance between the center of mass of two stars needed for a collision, i.e. the sum of the radii of the colliding stars, 
$\overline{m}_l$ is the average stellar mass in cell $l$, 
$t_{\mathrm{coll},l}$ is the collision timescale in cell $l$, and 
the factor $1/2$ is introduced to prevent double counting due to the fact that two stars participate in the collisions. In practice, we assume that the surrounding star $i$ collides in the simulation with probability 
\begin{align}
\label{eq:p_coll}
P_{\mathrm{coll},i} = 2 \sqrt{\pi} n_{*,l} \sigma_{*,l} 
\left( 
 R_{\mathrm{coll},i}^2+\frac{Gm_i}{\sigma_{*,l}^2}R_{\mathrm{coll},i}
\right)
\Delta t ,
\end{align}
during a time step, where the collision radius is approximated by $R_{\mathrm{coll},i}=2R_i$. 
For describing collisions between two surrounding stars $i$ and $j$, 
we assume that the relative velocity $v_\mathrm{rel,*}$ is drawn from a Maxwellian distribution with the dispersion of $\sqrt{2}\sigma_*$ as given in Eq.~(8.45) in \citet{Binney08}.  
Collisions may occur when the number of surrounding stars within a cell is more than one.

For collisions among contracted stars, when this relative velocity $v_\mathrm{rel,*}$ exceeds the escape velocity from the stars, $v_\mathrm{esc}=[2G(m_i+m_{j})/(R_i+R_j)]^{1/2}$, contracted stars lose a significant amount of mass at collision instead of simply coalescing into one remnant star \citep{Freitag05}. 
For simplicity, we assume that when $v_\mathrm{rel,*}>v_\mathrm{esc}$ for contracted stars, the colliding surrounding stars are completely disrupted. 
However, the fraction of the released gas mass that accretes onto the central star and that is converted to form new stars is not well understood. 
In this study, the mass released during collisions is added to the inflowing gas ${\dot M}_\mathrm{in}$ from large scales (equation~\ref{eq:Mdotin}). The inflowing gas is mostly converted to new surrounding stars during the early phase of the evolution (see \S~\ref{sec:star_formation}) and it is mostly accreted onto the central star when ${\dot M}_\mathrm{in}\sim {\dot m}_\mathrm{cent}$ (see \S~\ref{sec:accretion}). 
For collisions with $v_\mathrm{rel,*}<v_\mathrm{esc}$ between contracted surrounding stars, we assume that the stars coalesce without any mass loss. 
When two stars $i$ and $j$ coalesce, we assume that $m_j$ accretes onto $m_i$. The merger remnant star may either become an expanding star or a contracted star depending on the time-smoothed accretion rate as defined in \S \ref{sec:stellar_evolution}.

When surrounding stars are in an expanding phase (conditions specified in \S~\ref{sec:star_formation}), collisions become more frequent  
\citep{Boekholt18}. The mass loss during such collisions is also significantly different from that of contracted stars. 
\citet{AlisterSeguel19} have recently investigated the effect of mass loss during collisions on mass growth, relevant to collisions between contracted stars, in which high mass-loss rates are predicted.

We used the fraction of total mass lost during collisions among expanding stars from figure 8 of \citet{Adams04} as
\begin{equation}
\label{eq:mass_loss}
f_\mathrm{loss}=f_\mathrm{loss,max}~
10^{-1.2\left(\frac{R_\mathrm{p}}{R_{\mathrm{coll},i}} \right)}, 
\end{equation}
where $R_\mathrm{p}$ is the pericenter distance at collision. 
Referring to \citet{Adams04}, we set $f_\mathrm{loss,max}=0.16$ as a fiducial value, 
which is roughly consistent with the results by \citet{Bailey99}. 
Note that since \citet{Adams04} simulated collisions between an expanding star and a contracted star, $f_\mathrm{loss}$ for collisions between expanding stars may become lower than that in equation (\ref{eq:mass_loss}). 
To see the effect of $f_\mathrm{loss,max}$ on our results, we compared the final mass of the central star in one of the models (``Model 2'' below) with $f_\mathrm{loss,max}=0.16$, 0.3, and 1. 
The final mass of the central star was found to be $3.3\times 10^3~\Msun$, $2.6\times 10^3~\Msun$, and $1.8\times 10^3~\Msun$, respectively, and the total mass lost at collisions was $1.6\times 10^3~\Msun$, $2.9\times 10^3~\Msun$, and $4.0\times 10^3~\Msun$, respectively. 
Thus the final mass of the central star is affected by at most a factor $\sim 2$ due to $f_\mathrm{loss,max}$. 

The pericenter distance is related to the impact parameter $b$ through \citep[e.g.][]{OLeary09}
\begin{equation}
R_\mathrm{p} = \left(\sqrt{\frac{1}{b^2}+\frac{G^2(m_i+m_{j})^2}{b^4 v_\mathrm{rel,*}^4}} + \frac{G(m_i+m_{j})}{b^2 v_\mathrm{rel,*}^2}\right)^{-1}. 
\end{equation}
We set the distribution of $R_{\rm p}$ so that $b^2$ is uniformly distributed between 0 and $b_{\rm max}^2$, 
where $b_{\rm max}$ is the maximum impact parameter at which collision occurs ($b=b_{\rm max}$ at $R_{\rm p}=R_{\rm coll}$). 
The fraction of mass $f_{\rm loss}$ (Eq.~\ref{eq:mass_loss}) is subtracted from the mass of the collided surrounding stars, and 
added to the inflow rate ${\dot M}_\mathrm{in}$. 
Following \citet{Bailey99}, we also set the condition for merger into a single star to 
\begin{equation}
\frac{R_\mathrm{p}}{R_\mathrm{coll}} < -0.75 \left(\frac{v_\mathrm{rel,*}}{v_\mathrm{esc}}\right)+1.3.
\end{equation}
Even when the colliding surrounding stars merge into one single star, the collision leads to some mass loss in the case where at least one of the two colliding stars is expanding, according to equation (\ref{eq:mass_loss}).

\subsection{Stellar and gaseous accretion onto the central star}

\subsubsection{Stellar accretion}

Surrounding stars collide with and accrete onto the central star when the distance from the central star to some surrounding star $r_{i}$ becomes smaller than the sum of the radii $R_\mathrm{cent}+R_i$.
After a star accretes onto the central star, we add the mass of the accreted surrounding star to the mass of the central star, and the radius of the central star increases according to equations (\ref{eq:r_expand}) or (\ref{eq:r_contract}). 
We assume no mass loss during this collision/accretion event. 
\citet{Freitag05} show that when the collision velocity ($v_\mathrm{coll}$) is smaller than the escape velocity 
from the surface of the
collided star ($v_\mathrm{esc}$), the mass loss is small. 
If the accreted surrounding star orbits in a gravitational potential dominated by the central star, the collision velocity becomes smaller than the escape velocity from the central star. 
This can be violated and 
some fraction of the envelope of the central star will be lost 
if surrounding stars accrete on highly eccentric orbits, which cannot be accounted for in our present model. 
After accreting a surrounding star,  
we assume that the envelope of the central star is heated since the orbital energy of the accreted surrounding star is converted to thermal energy in the envelope of the central star. 
The accreted surrounding star then sinks to the core of the central star, and the central star is expected to expand, similar to the case for gas accretion \citep{Sakurai15}. 
We determine the expansion rate of the central star according to the averaged mass accretion rate (\S \ref{sec:stellar_evolution}).

\subsubsection{Gas accretion}
\label{sec:accretion}

\citet{Inayoshi18} have considered radiatively inefficient accretion onto a compact object and generalized Bondi accretion to a case with angular momentum. They have found that when the angular momentum is low, so that the centrifugal radius is well inside the Bondi radius and a compact accretion disk forms around the central object, the accretion rate ${\dot m}_{\mathrm{acc},i}$ onto the central object (in our case the $i^{\rm th}$ star) is given by 
\begin{equation}
\label{eq:accretion}
{\dot m}_{\mathrm{acc},i}= f_\mathrm{sup}
{\dot m}_{\mathrm{BHL},i}.
\end{equation}
where 
\begin{equation}
f_\mathrm{sup}=\min
\left\{
 1,\max
 \left[
   \left(
     \frac{\alpha_\mathrm{SS}}{0.01}
   \right)^{0.62} 
   \frac{r_{\mathrm{in},i}}{R_{\mathrm{BHL},i}},f_\mathrm{sup,min}
 \right]
\right\}
\end{equation}
is the suppression rate from the Bondi accretion rate, 
$\alpha_\mathrm{SS}$ is the viscosity parameter in the standard thin $\alpha$-disk model~\citep{Shakura73}, 
${\dot m}_{\mathrm{BHL},i}=4\pi G^2 n_\mathrm{gas} \mu m_\mathrm{H} m_i^2/(c_s^2+v_i^2)^{3/2}$ is the Bondi-Hoyle-Lyttleton accretion rate,
$R_{\mathrm{BHL},i}=Gm_i/(c_s^2+v_i^2)$ is the Bondi-Hoyle-Lyttleton radius, 
$r_{\mathrm{in},i}$ is the inner radius, and $f_\mathrm{sup,min}$ is the minimum of the suppression rate. 
\citet{Inayoshi18} found that $f_\mathrm{sup,min}\sim 10^{-2}-10^{-3}$. 
We set $f_\mathrm{sup,min}=0.003$. 
The inner radius is the inner boundary of the calculation introduced in \citet{Inayoshi18} due to the computational limit. 
The inner radius is considered to correspond to the stellar radius 
$r_{\mathrm{in},i}=R_i$. 
We set $\alpha_\mathrm{SS}=0.01$ as a fiducial value. This value is motivated by the results for a weak vertical magnetic field by \citet{Bai13}, which simulates the magnetorotational instability turbulence (however see also \citealt{King+2007}).  
We limit the maximum accretion rate to ${\dot m}_{\mathrm{acc},i}={\dot m}_{\mathrm{BHL},i}$, if ${\dot m}_{\mathrm{acc},i}>{\dot m}_{\mathrm{BHL},i}$ given by Eq.~\eqref{eq:accretion}, since in this case equation (\ref{eq:accretion}) becomes invalid, which describes the reduction in the accretion rate relative to the Bondi-Hoyle-Lyttleton rate due to rotation. 
We do not consider the enhancement of the gas density due to the $N$-body accretion \citep{Kaaz19}, 
since the upper limit on the density of gas outside of the Bondi-Hoyle-Lyttleton radius is given by $n_\mathrm{c}$. 

If the velocity of the $i^{\rm th}$ surrounding star is sufficiently high, $R_{\mathrm{BHL},i}$ may become smaller than $R_i$. In this case, the gas accretion rate is determined by direct collision with the stellar surface, 
${\dot m}_{\mathrm{coll},i}=\pi R_i^2 n_\mathrm{gas} \mu m_\mathrm{H} v_i$. 
We set the accretion rate to $\mathrm{max}[{\dot m}_{\mathrm{acc},i},{\dot m}_{\mathrm{coll},i}]$. 
Furthermore, when the total accretion rate $\Sigma_i{\dot m}_{i}$ onto all stars exceeds the inflow rate from large scales ${\dot M}_\mathrm{in}$, we normalize the respective accretion rate of each star by the inflow rate by multiplying it by ${\dot M}_\mathrm{in}/\Sigma_i{\dot m}_{i}$. 
When $\Sigma_i{\dot m}_{i}\sim{\dot M}_\mathrm{in}$, the gas density should be depleted.  However, for simplicity, we assume that the gas density distribution is unchanged; this is justified since whenever this condition is satisfied, the dynamical evolution of surrounding stars is hardly affected by the presence of gas because the gravitational potential is dominated by stars in later phases and star formation ceases.

In cases in which the Bondi mass $M_{\mathrm{Bondi},i}=\frac{4}{3}\pi R_{\mathrm{BHL},i}^3 n_\mathrm{gas}\mu m_\mathrm{H} $, i.e. the gas mass within the Bondi-Hoyle-Lyttleton radius ($R_{\mathrm{BHL},i}$) of the $i^{\rm th}$ star, is larger than the stellar mass $m_i$, the gas within the Bondi-Hoyle-Lyttleton radius can be unstable to fragmentation 
since the Jeans instability can be significant in weak shear regions \citep[e.g.][]{Elmegreen94,KimOstriker01,KimOstriker02}. 
If fragmentation is significant, it is not obvious what fraction of the gas can accrete onto the star. 
The fraction depends on cooling, turbulence \citep[e.g.][]{Clark11,Greif11,Elmegreen11,Dopcke13}, 
and the efficiency of angular momentum transfer \citep[e.g.][]{Thompson05}. 
Following the prescription in \citet{Ryu16}, when the Bondi mass exceeds the stellar mass $M_{\mathrm{Bondi},i}>m_i$, we reduce the accretion rate $\dot m_i$ by a constant factor $f_{\rm red}$.
We assume $f_\mathrm{red}=10^{-3}$ as a fiducial value. 
Even when fragmentation is expected within $r_{\mathrm{BHL},i}$, we assume that surrounding stars form at $r_\mathrm{c}$ (\S \ref{sec:star_formation}). 

 Thus, in summary we calculate the gas accretion rate of stars as
\begin{align}
    \dot{m}_{i} = 
    \left\{
    \begin{array}{cc}
        {\dot m}_{i,0} &  \mathrm{if}~ M_{\mathrm{Bondi},i} < m_{i}, \\
        f_\mathrm{red} {\dot m}_{i,0} & \mathrm{if}~ M_{\mathrm{Bondi},i} \geq m_{i},
    \end{array}
    \right.
\end{align}
where
\begin{align}
    \dot{m}_{i,0} &= 
    \min\left(
        \dot{m}_{i,1}, 
        \frac{\dot{m}_{i,1}\dot{M}_\mathrm{in}}{\sum_i \dot{m}_{i,1}} \right)\,,
 \end{align}
 and 
\begin{align}
    \dot{m}_{i,1} &= \max({\dot m}_{\mathrm{acc},i},{\dot m}_{\mathrm{coll},i})\,.
\end{align}

\subsection{Feedback effects}
\label{sec:gas_ejection}

Photoionization and supernova feedback play key roles in ejecting gas from pre-galactic halos \citep{Whalen+2004,Kitayama04,Kitayama05}. We did not take into account feedback from supernova explosions since our simulations are limited to the time until a first supernova explosion at $3$~Myr. 
\citet{Kitayama04} have shown that when the production rate of ionizing photons ${Q}_\mathrm{ion}=\sum_i {Q_{\mathrm{ion},i}}$ is below the critical value ${Q}_\mathrm{cri}\sim 10^{51}\,\mathrm{s^{-1}}(M_\mathrm{halo}/10^7\,\Msun)^{8/5}[(1+z)/15]^{12/5}$, the gas density is not affected by photoionization feedback. On the other hand, when ${Q}_\mathrm{ion}$ exceeds ${Q}_\mathrm{cri}$, gas is blown away from the halo. We adopt this criterion as the gas ejection condition.
${Q_{\mathrm{ion},i}}$ strongly depends on whether the $i^{\rm th}$ star is in the expanding phase or the contracting phase, with ${Q_{\mathrm{ion},i}}\sim 10^{36}~\mathrm{s^{-1}} (m_i/10M_\odot)^2$ and $\sim 10^{48}~\mathrm{s^{-1}} (m_i/10M_\odot)^2$ in these phases, respectively \citep{Hosokawa12}. 
If ${Q}_\mathrm{ion}>{Q}_\mathrm{cri}$ is ever satisfied, all gas is assumed to be ejected from the system.

Although we add up the ionizing photons emitted by low-mass stars ($\lesssim 1\,\Msun$), 
these photons do not affect bulk gas dispersal due to their low numbers and because they are trapped within their parent stars' Bondi radii. Furthermore, the contraction timescale for low-mass stars exceeds the calculation time (3 Myr), and we expect that low-mass stars  contribute negligibly to the total photon emission rate. 
Indeed, we find below that gas dispersal (when it occurs) is caused by the contraction of the central star in our models.

After the gas is ejected, gas accretion, star formation, and gas dynamical friction are all assumed to stop operating 
(${\dot m}_i={\dot m}_\mathrm{SF}=n_\mathrm{gas}(r)=0$ for all $i$ and $r$) during the rest of the simulation.\footnote{Note that even if gas is released during collisions among surrounding stars after this point,  we assume that it is also blown away by feedback in this phase.}
The radial position of surrounding star $i$ increases to $r_i+\Delta r$ due to the decrease of the potential energy as 
\begin{align}
\Phi(r_i+\Delta r)-\Phi_\mathrm{gas}(r_i+\Delta r)+k_\mathrm{ej}(r_i+\Delta r)\nonumber\\%
=\Phi(r_i)-\Phi_\mathrm{gas}(r_i)+k_\mathrm{be}(r_i), 
\end{align}
where $k_\mathrm{be}(r)$ and $k_\mathrm{ej}(r)$ are the specific kinetic energies of an object at radius $r$ with and without gas, respectively. 
As in equation (\ref{eq:migration_rate}), we use the zero-eccentricity approximation when we calculate the change in the radial position of surrounding stars. 

Although we assume that gas is ejected instantly, 
the ejection timescale is roughly given by the size of the gas cloud over the ejection speed. 
In our models, the gas distribution affects the dynamical evolution of surrounding stars, and most surrounding stars are distributed within 0.1 pc. The ejection timescale for gas within 0.1 pc is $\sim 10^4$ yr when the ejection speed is $\sim 10\,\mathrm{km/s}$, which is set to a rough value of the sound speed of ionized gas. 
Thus the ejection timescale is much smaller than our total calculation timescale of 3~Myr, which justifies the assumption of instant gas ejection.

\begin{figure*}
\begin{center}
\includegraphics[width=180mm]{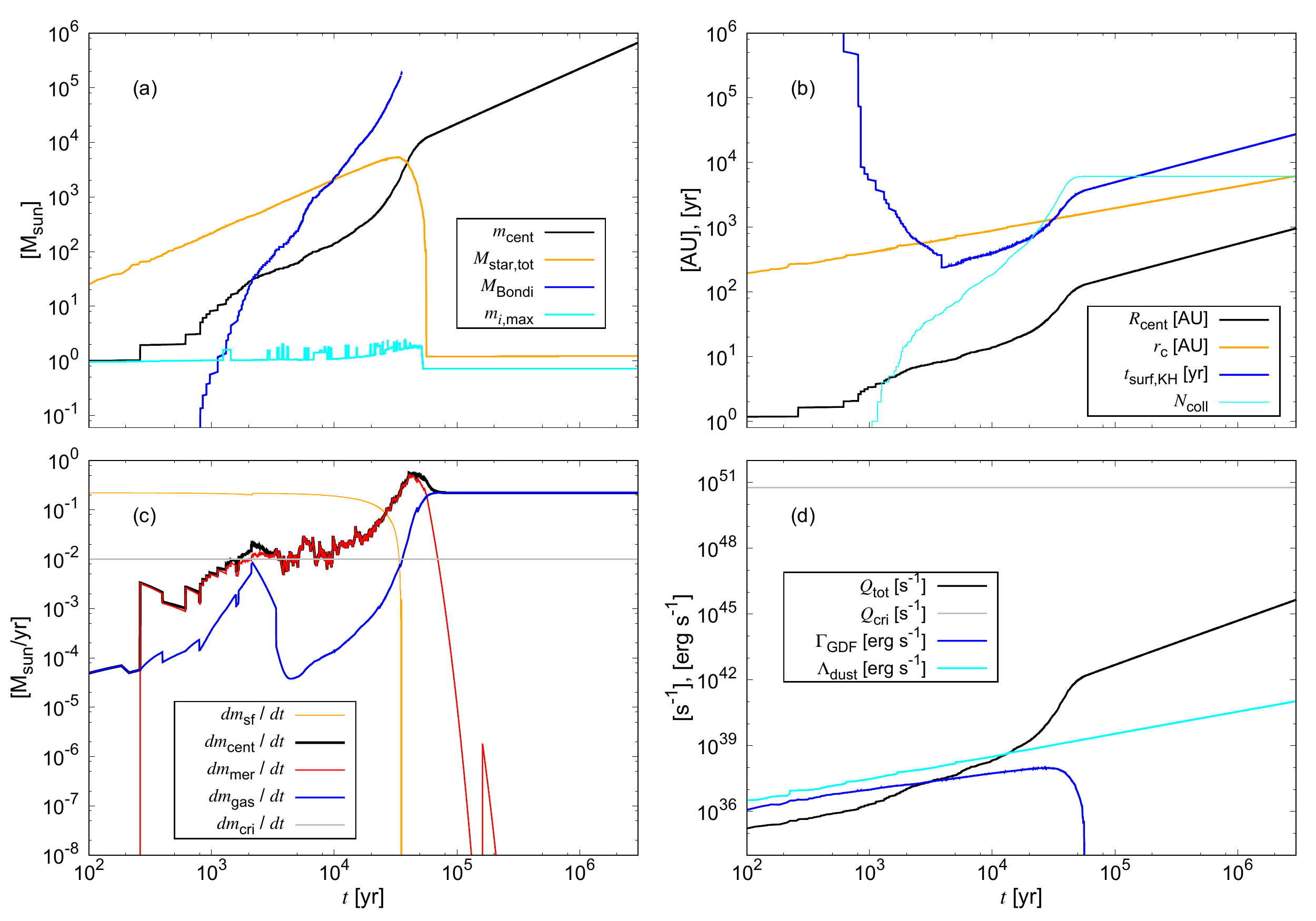}
\caption{
Evolution of several quantities in the fiducial Model 1. 
(a) The mass of the central star (black), the total mass of surrounding stars (orange), the total gas mass 
within the Bondi radius of the central star 
(blue), and the most massive star among surrounding stars (cyan). 
(b) The radius of the central star (black), the core radius of collapsing gas (orange), the Kelvin-Helmholtz (KH) timescale for the stellar surface of the central star, $t_\mathrm{surf,KH}$ (blue), and the number of collisions among surrounding stars (cyan). 
(c) The growth rate of the central star (black), the rate of stellar accretion onto the central star (red), the gas accretion rate onto the central star (blue), 
the total star formation rate (orange), and the critical accretion rate below which the central star contracts when the age of the central star exceeds the KH timescale $t_{\mathrm{KH}}$ (gray).
The black, red, and blue lines are smoothed on a timescale of $t_{\mathrm{surf,KH}}$ since the behavior (contraction or expansion) of the central star depends on the growth rate averaged over this timescale. 
(d) Black and gray lines are the total production rate of ionizing photons and the critical production rate of ionizing photons at which gas is ejected from the halo, respectively. 
The cyan and blue lines show the cooling rate by dust grains and the heating rate due to gas dynamical friction by surrounding stars, respectively. 
In this model, the high growth rate of the central star enables it to continue expanding and growing into a supermassive star with $6.7\times 10^5\,\Msun$ at the end of the simulation at 3~Myr. 
}
\label{fig:cent_ev_fid}
\end{center}
\end{figure*}

\begin{figure*}
\begin{center}
\includegraphics[width=180mm]{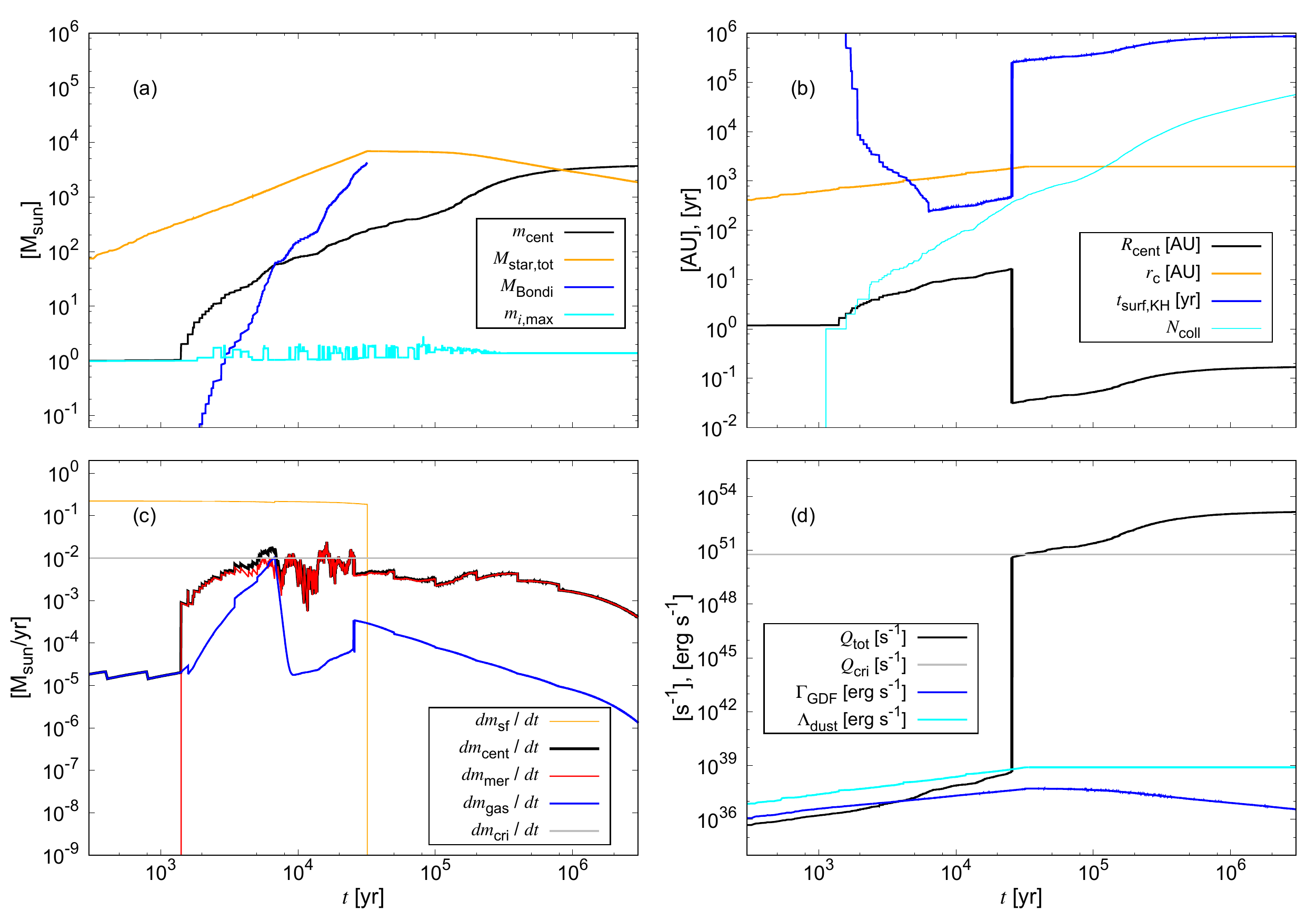}
\caption{
Same as Fig.~\ref{fig:cent_ev_fid}, but with a reduced $n_\mathrm{c}=3\times 10^{10}\,\mathrm{cm^{-3}}$ (Model 2 in Table~\ref{table_results}), illustrating a case when the central star contracts. 
The slow decline of the gas accretion rate onto the central star (blue line in panel (c)) after gas dispersal is due to smoothing in time calculated from Eq.~\eqref{eq:time_smoothing}. 
}
\label{fig:cent_ev_cont}
\end{center}
\end{figure*}

\begin{figure}
\begin{center}
\includegraphics[width=85mm]{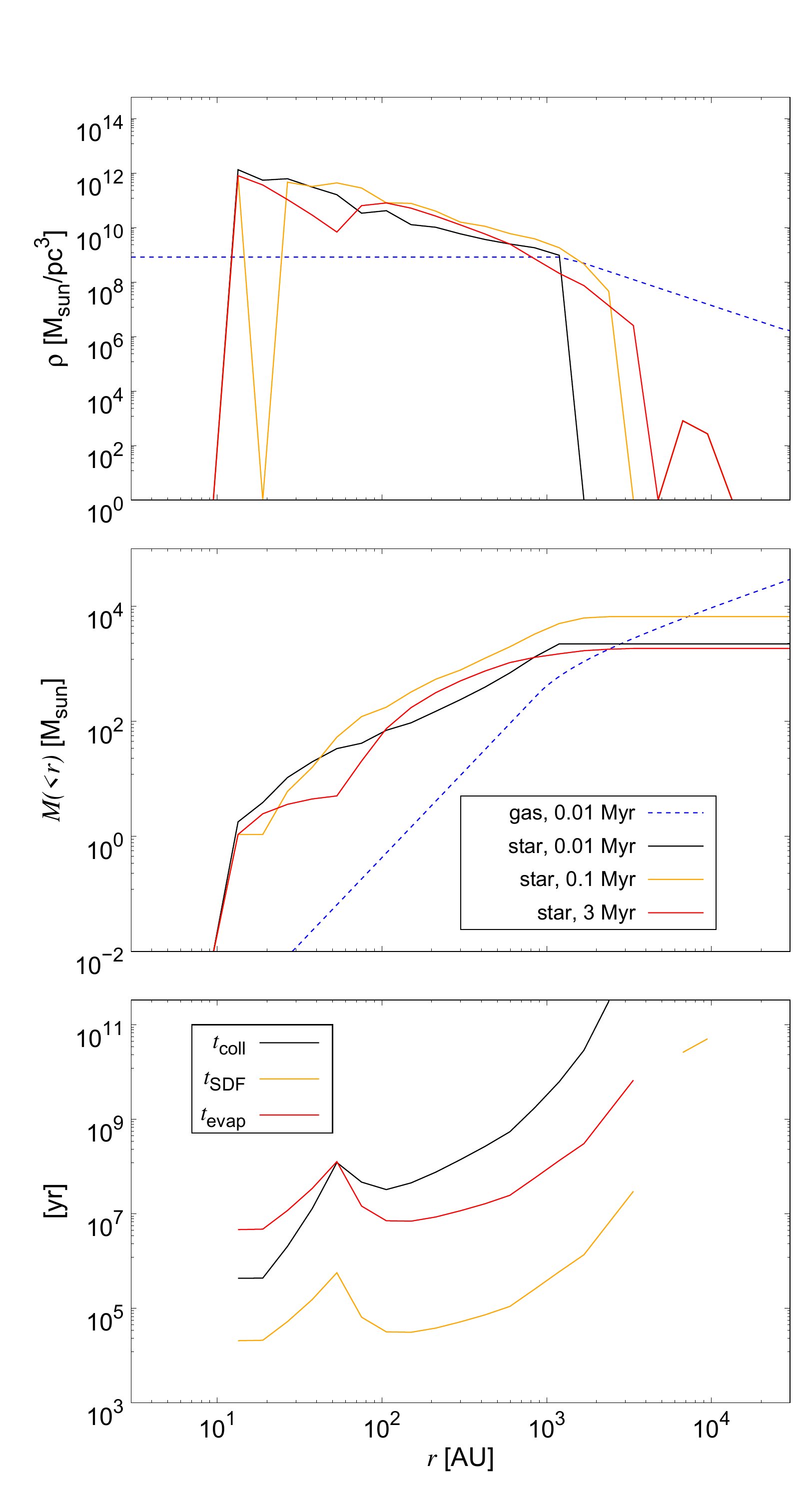}
\caption{
Snapshots of the density and mass profiles and related timescales in Model~2. 
Upper panel: gas and stellar density profiles. 
The lines show the gas density at 0.01~Myr (dashed blue), and
the stellar densities at 0.01~Myr (black), 0.1~Myr (orange), and 3~Myr (red), respectively. 
Middle panel: enclosed mass profile for gas and surrounding stars. 
The lines show the enclosed mass profile for gas at 0.01~Myr (dashed blue), and
for surrounding stars at 0.01~Myr (black), 0.1~Myr (orange), and 3~Myr (red), respectively. 
Lower panel: timescales for collision (black), 
stellar dynamical friction (orange), 
and evaporation (red) 
for the stellar distribution at 3~Myr. 
}
\label{fig:snapshot}
\end{center}
\end{figure}

\begin{figure}
\begin{center}
\includegraphics[width=85mm]{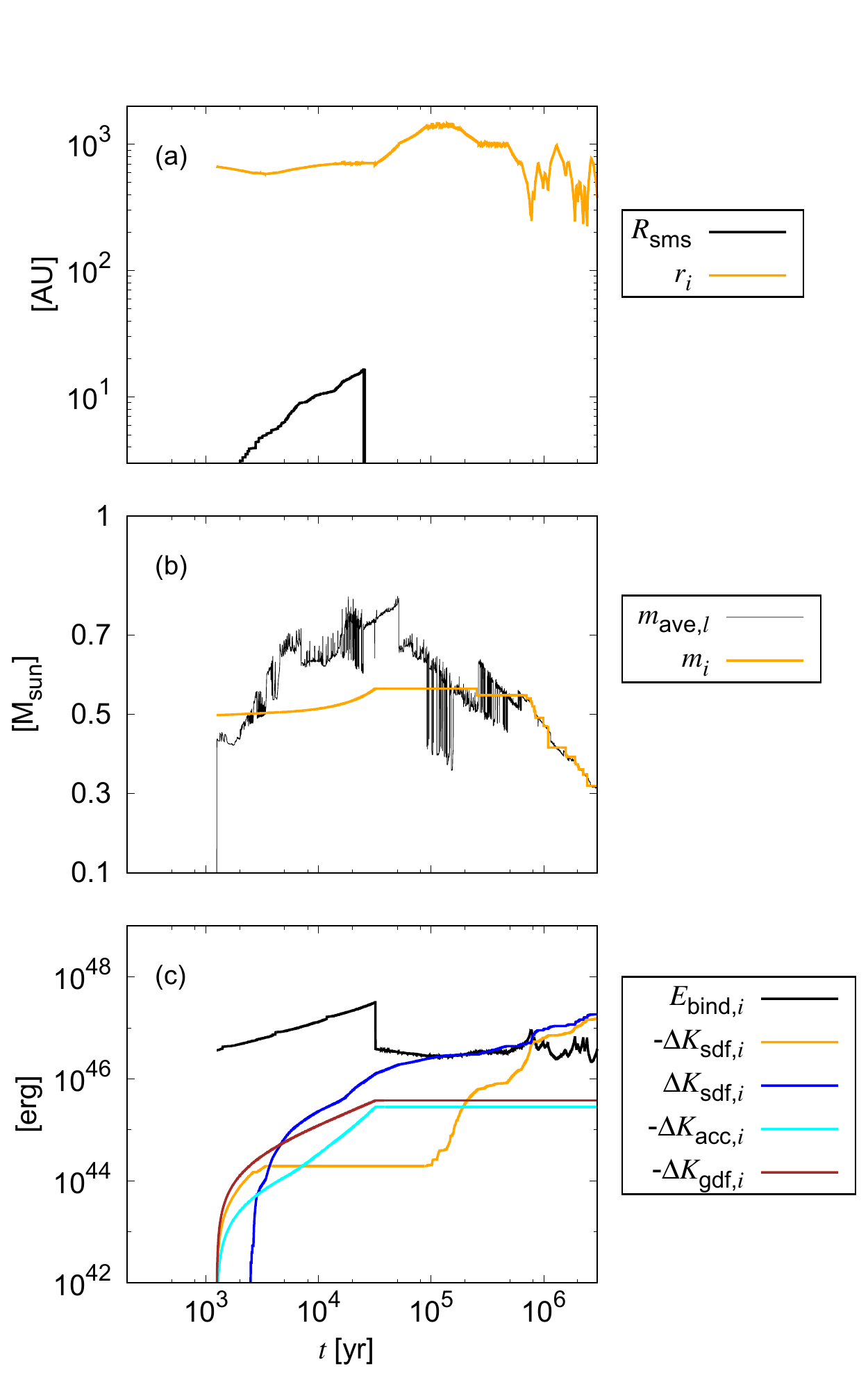}
\caption{
The evolution of one of the surrounding stars, born with an initial mass of $m_i=0.50\,\Msun$ formed at $t=1.2\times 10^3$ yr in Model 2. 
(a) The radial position (orange) and the radius of the central star (black). 
(b) The mass of the star (orange) and 
the average mass of surrounding stars in the cell $l$ hosting the star $\overline{m}_l$ (black). 
(c) The binding energy of the star (black) and 
the cumulative change in the kinetic energy due to the stellar dynamical friction (orange and blue), 
accretion drag (cyan), 
and gas dynamical friction (brown). 
Decrease and increase of the kinetic energy by stellar dynamical friction 
are shown separately by orange and blue lines, respectively. 
This star decreases its mass due to frequent collisions, and ends up as a less massive star orbiting at $\sim 380$AU at 3~Myr.
}
\label{fig:ev_star_osc}
\end{center}
\end{figure}

\section{Results}
\label{sec:results}

\subsection{Central star evolution}
\label{sec:central_evolution}

We have performed several numerical calculations using the above semianalytical model.
We first present the evolution of the central star in the fiducial model (labeled as Model 1 in Table~\ref{table_results}). Figure~\ref{fig:cent_ev_fid} shows the evolution of several other quantities in this model. 

In the early stages, the gas accretion rate onto the central star ${\dot m}_{\mathrm{acc,}i}$ is as low as  $\sim 6\times 10^{-5}(m_i/1\,{\rm M_\odot})^{1.5}\,{\rm M_\odot}/\mathrm{yr}$ (blue line in panel (c) of Figure \ref{fig:cent_ev_fid}), and almost all of the gas from flowing in from large scales is converted into new surrounding stars, at the rate of ${\dot m_\mathrm{SF}}\sim{\dot M_\mathrm{in}}\simeq 0.22\,{\rm M_\odot}/\mathrm{yr}$ (orange line in panel (c)), and so a dense $\sim 10^3\,{\rm M_\odot}$ stellar cluster forms in $\sim 10^4$ years (orange line in panel (a)).

At $2.6\times 10^2$ yr, the central star
accretes the first surrounding star (red line in panel (c)). The accretion rate of surrounding stars subsequently gradually increases, due to the increasing radius of the central star (eventually to $\sim 10^3$ AU) 
as well as due to the increase in the number of surrounding stars (red line in panel (c), black line in panel (b), and orange line in panel (a)). At the same time, the surface KH time $t_\mathrm{surf,KH}$ for the central star decreases significantly (blue line in panel (b)) due to its increase in mass, which significantly raises the luminosity up to  $m_\mathrm{cent}\sim6\,\Msun$ \citep{Hosokawa12}.

Before the radius of the central star increases to $10$ AU, 
77 collisions occur between surrounding stars (cyan line in panel (b)). All of these collisions occur between expanding surrounding stars, $4.7\,\Msun$ is lost during stellar collisions, 
and five pairs of them merge as a result of collisions. 
The mass lost during collisions is added to the gas inflow rate ${\dot M}_\mathrm{in}$.

At $1.7\times  10^3$ yr the central star's mass is $m_\mathrm{cent}=19\,\Msun$, and its growth rate exceeds the critical rate required to inhibit contraction (${\dot m}_\mathrm{cri}=0.01\,\mathrm{\Msun/yr}$ see \S~\ref{sec:stellar_evolution}; black and gray lines in panel (c)). 
The accretion rate of surrounding stars onto the central star subsequently
increases further, due to the increasing radius of the central star, the increased total number of surrounding stars in the cluster, and the increased masses of the surrounding stars
(black line in panel (b), orange and cyan lines in panel (a)). Thus the mass of the central star rapidly increases by stellar bombardment in this phase. 

At $t = 2.2\times 10^3\,\mathrm{yr}$, the Bondi mass of the central star exceeds its own mass, $m_\mathrm{cent}=31\,\Msun$ (blue and black lines in panel (a)), and the gas accretion rate is reduced by $f_\mathrm{red}=10^{-3}$ due to the fragmentation of gas within the Bondi-Hoyle-Lyttleton radius (see \S~\ref{sec:accretion}).

Since the cooling rate due to emission by dust grains (equation~\ref{eq:cooling_dust}) 
always exceeds the heating rate due to gas dynamical friction (blue and cyan lines in panel (d)), surrounding stars continue forming at the core radius (\S \ref{sec:star_formation}).

At $\gtrsim 10^5$ yr, gas accretion begins to dominate the central star's growth rate. In this phase, most of the gas flowing in from large scales ${\dot M}_\mathrm{in}$ is accreted onto the central star. 
Star formation ceases, and a large number of surrounding stars are absorbed by the central star due to the radial growth of the central star (orange lines in panels (c) and (a) and black line in panel (b)).  Stellar bombardment keeps the central star in the bloated state. Hence, the central star continues growing, and reaches $\gtrsim 10^5\,\Msun$ without any contraction.

During the evolution, $6.1\times 10^3$ collisions occur (cyan line in panel (b)), all are between expanding surrounding stars, and 31 pairs of them merge as a result of collisions. In total $3.5\times 10^2\,\Msun$ mass is lost by surrounding stars during collisions (\S \ref{sec:mergers}). Thus most collisions between surrounding stars result in a relatively small amount of mass loss in 
the fiducial model.

To clarify the importance of each mechanism for the growth of the central star, we repeat the above calculation in several variants of the fiducial model, in which parameter settings remain unchanged but some mechanism is turned off and does not operate. 

To check whether the stellar bombardment plays an important role, we first run the model with the fiducial settings but no migrating motion for surrounding stars. 
In this model, the final mass of the central star is found to be $32\,\Msun$. Thus via gas accretion alone, we find that the central star contracts and cannot grow into an SMS. 

We next investigate the importance of dynamical friction.  With stellar dynamical friction turned off, the final mass of the central star is $750\,\Msun$.  On the other hand, in the model without gas dynamical friction or without gas accretion drag,
respectively, the final masses of the central stars are $6.7\times 10^5\,\Msun$ and $6.6\times 10^5$. 
We conclude that the migration of surrounding stars is dominated by stellar dynamical friction rather than gas dynamical friction and gas accretion drag. 
This is essentially because the density of stars dominates the density of gas. For example, in Model 1,  the gas mass within the core radius of $r_\mathrm{c}=880$ AU at $t=10^4$ yr is $9.9\times 10^2\,\Msun$, while that of stars is $2.1 \times 10^3\,\Msun$. The stellar density increases closer to the SMBH, while the gas density is set to be constant within the core (e.g. upper and middle panels of Figure~\ref{fig:snapshot}).

We further simulate a model with the fiducial settings, but without allowing the stars to expand, and instead always setting their radii to the value in Eq.~\eqref{eq:r_contract}. In this model, the rate of stellar accretion onto the central star does not increase beyond $\sim 0.005\,\Msun/\mathrm{yr}$, and the final mass is found to be $1.7\times 10^3\,\Msun$. Thus the bloating of stars is required to facilitate the growth of the central star due to stellar accretion.

We next present a case in which the central star contracts before it collapses to a BH. 
Figure~\ref{fig:cent_ev_cont} shows the results in Model~2, which differs from Model~1 only by a modified value of the gas density (reduced by a factor of 3 to $n_\mathrm{c}=3 \times 10^{10}\,\mathrm{cm^{-3}}$). 
Initially the radial position at which surrounding stars form is about a factor of 1.5 larger than in Model~1 (orange lines in panels (b) in Figures~\ref{fig:cent_ev_cont}~versus~\ref{fig:cent_ev_fid}). 
Stellar dynamical friction becomes less efficient due to the lower stellar density, 
and the average accretion rate of surrounding stars onto the central star becomes lower than in Model~1 (red line in panel (c) of Figure~\ref{fig:cent_ev_cont}). 
At $t_\mathrm{KH,ZAMS} (\sim 3\times 10^4$ yr) for the central star, 
the central star contracts, and then the production rate of ionizing photons exceeds the critical value for gas ejection from the system (black and gray lines in panel (d)). 
After the ejection, gas accretion and star formation cease (blue and orange lines in panel (c)), and the rate of accretion of surrounding stars decreases (red line in panel (c)). 
The radius of a star is predicted to contract in $\sim 10^2-10^3$ yr  \citep[e.g.][]{Sakurai15}, which justifies the assumption of abrupt contraction in our calculations. 
The central star continues to grow by accreting surrounding stars, but at a more moderate rate, reaching the mass of $3.7\times 10^3\,\Msun$ at 3~Myr.  Therefore a massive BH may still be produced in Model~2, but the mass of this massive BH is $\approx 100$ times below that of the BH remnant in Model 1.

The top panel of Figure \ref{fig:snapshot} shows the stellar and gas density profiles for Model 2. 
The power-law slope of the stellar density is almost unchanged during the evolution (black, orange, and red curves in the top panel). Coincidentally, such self-similar evolution is also expected for the core collapse of a self-gravitating system driven by two-body relaxation \citep{Binney08}. The evolution of the stellar density in our model is driven by the combination of gas and stellar dynamical friction, Bondi accretion, and star formation. 
On the other hand, the outer cutoff of the stellar density distribution slowly evolves from 0.1~to 3~Myr (orange and red curves in the top panel of Figure \ref{fig:snapshot}). 
This is because the timescale for stellar dynamical friction ($t_\mathrm{SDF}\equiv v_\mathrm{Kep}(r)/a_{\mathrm{SDF},i}$ with ${m}_i=\overline{m}_l$) at this position (orange curve in the bottom panel) exceeds the calculation timescale of 3~Myr. 
At 3~Myr, the density profile contains 
zero surrounding stars within $10$AU, and $\approx$50 surrounding stars within 100 AU.

\subsection{Evolution of surrounding stars}
\label{sec:stellar_evolution_result}

Figure~\ref{fig:ev_star_osc} shows the evolution of one of the surrounding stars in Model 2. This star is born at $1.2 \times 10^3$ yr with mass $m_i=0.50\,\Msun$ at $r_{i}=660$ AU. In the early phase, due to gas dynamical friction, accretion drag and stellar dynamical friction, 
the star migrates inwards very slightly (orange line in panel (a) and brown, cyan, and orange lines in panel (c)). 

At $\gtrsim 2.5 \times 10^3$ yr, the average mass in the cell hosting this star becomes more massive than the mass of the star due to the formation of new stars within the cell (orange and black lines in panel (b)). The star therefore begins to migrate outward due to mass segregation.

At $3.2\times 10^4$ yr, gas is ejected from the system due to photoionization feedback (as was shown by the gray and black lines in panel (d) of Fig.~\ref{fig:cent_ev_cont}), and so the binding energy of this star decreases abruptly (black line in panel (c) in Fig.~\ref{fig:ev_star_osc}). Gas dynamical friction and gas accretion stop operating due to the lack of gas around the star. 
Since star formation also stops operating and massive surrounding stars migrate inward, the average masses in cells at $\sim 100-1000$ AU begin to decrease. 
At $\sim 2\times 10^5$ yr, this star also begins to migrate inward.
In the inner regions, the star and other surrounding stars lose some fraction of their mass due to frequent collisions (black and orange lines in panel (b)). 
Since the direction of migration due to stellar dynamical friction is influenced by the mass of the star compared to that of surrounding stars,  
the star wanders around $\sim 500$ AU. 
When the central star collapses into a massive BH at $3\times 10^6$ yr, this star orbits at $r_i= 380$ AU, 
and the mass is $m_i=0.32\,\Msun$. 
Hence surrounding stars are redistributed mainly by mass segregation driven by stellar dynamical friction, and only more massive stars can migrate toward the central star. 
In later phases, frequent collisions reduce the masses of surrounding stars, and they prevent accretion of surrounding stars onto the central star.

\begin{table*}
	\caption{
		The results of our simulations in the fiducial model (Model~1) and 21 variants.
        The columns show several input and output parameters in each case, as follows: the model number, 
        the core gas number density ($n_\mathrm{c}$), 
        the gas temperature ($T$), 
        the reduction factor for the gas accretion rate when the Bondi mass exceeds the central mass ($f_\mathrm{red}$), 
        the power-law index of the stellar IMF ($\beta$), 
        the maximum initial mass of stars ($m_\mathrm{0,max}$), 
          the mass of the central star at the end of the simulation at 3~Myr ($m_\mathrm{fin}$), 
           the total mass of surrounding stars accreted onto the central star ($m_\mathrm{acc,*}$), 
           the total mass lost during collisions ($M_\mathrm{loss}$), 
           the number of newly formed surrounding stars ($N_\mathrm{SF}$), 
        the number of surrounding stars accreted onto the central star ($N_\mathrm{acc}$), 
        the number of collisions between surrounding stars ($N_\mathrm{coll}$), 
        the mass of the central star and the time at the ejection of gas from the system ($m_\mathrm{ej}$ and $t_\mathrm{ej}$) for models in which such ejection occurs, 
        and the maximum value for the ratio of the heating rate by gas dynamical friction to the cooling rate by dust grains ($\gamma_\mathrm{HC}=\mathrm{max}(\Gamma_\mathrm{GDF}/\Lambda_\mathrm{dust}$)). 
}
\label{table_results}
\hspace{-15mm}
\begin{tabular}{c|c|c|c|c|c||c|c|c|c|c|c|c|c|c}

\hline
\multicolumn{6}{c}{Input} \vline& \multicolumn{9}{c}{Output}\\\hline
Model&$n_\mathrm{c}$
&$T/10^4$
&$f_\mathrm{red}$&$\beta$&
$m_\mathrm{0,max}$&
$m_\mathrm{fin}$&
$m_\mathrm{acc,*}$&
$M_\mathrm{loss}$&
$N_\mathrm{SF}$&$N_\mathrm{acc}$&
$N_\mathrm{coll}$&$m_\mathrm{ej}$&
$t_\mathrm{ej}$&
$\gamma_\mathrm{HC}$
\\
&
$[\mathrm{cm^{-3}}]$
&[K]
&&&
$[\Msun]$&
$[\Msun]$&
$[\Msun]$&
$[\Msun]$&
&&
&$[\Msun]$&
$[\mathrm{yr}]$&
\\\hline\hline

1&$10^{11}$&$1$&$10^{-3}$&0&1&
$6.7\times 10^5$&$9.6\times 10^3$&$3.5\times 10^2$&$8.7\times 10^3$&$8.7 \times  10^3$ &$6.1\times 10^3$&-&-&0.50\\\hline

2&$3\times 10^{10}$&$1$&$10^{-3}$&0&1&
$3.7\times 10^3$&$3.7\times 10^3$&$1.6\times 10^3$&$1.2\times 10^4$&$5.1 \times  10^3$ &$5.7\times 10^4$&$2.4\times 10^2$&$3.2\times 10^4$&0.19\\\hline

3&$10^{11}$&$0.5$&$10^{-3}$&0&1&
$2.4\times 10^5$&$3.8\times 10^3$&$1.6\times 10^2$&$4.1\times 10^3$&$4.1 \times  10^3$ &$3.3\times 10^3$&-&-&0.77\\\hline

4&$10^{11}$&$0.3$&$10^{-3}$&0&1&
$1.1\times 10^5$&$1.8\times 10^3$&$89$&$2.2\times 10^3$&$2.1 \times  10^3$ &$2.0\times 10^3$&-&-&0.67\\\hline

5&$10^{11}$&$0.1$&$10^{-3}$&0&1&
$2.8\times 10^2$&$2.4\times 10^2$&$57$&$5.0\times 10^2$&$3.2 \times  10^2$ &$3.3\times 10^3$&$2.4\times 10^2$&$4.8\times 10^4$&0.67\\\hline

6&$10^{11}$&$1$&$0$&0&1&
$6.7\times 10^5$&$6.7\times 10^5$&$4.3\times 10^2$&$6.7\times 10^5$&$6.7 \times  10^5$ &$7.6\times 10^3$&-&-&0.50\\\hline

7&$10^{11}$&$1$&$10^{-3}$&2.35&1&
$6.7\times 10^5$&$1.1\times 10^4$&$5.7\times 10^2$&$2.6\times 10^4$&$2.6 \times  10^4$ &$2.8\times 10^4$&-&-&0.18\\\hline

8&$3\times 10^{10}$&$0.5$&$10^{-3}$&0&1&
$1.7\times 10^3$&$1.7\times 10^3$&$6.2\times 10^2$&$4.8\times 10^3$&$2.4 \times  10^3$ &$2.4\times 10^4$&$2.4\times 10^2$&$3.5\times 10^4$&0.23\\\hline

9&$3\times 10^{10}$&$0.3$&$10^{-3}$&0&1&
$9.3\times 10^2$&$9.3\times 10^3$&$35$&$2.6\times 10^3$&$1.3 \times  10^3$ &$1.5\times 10^4$&$2.4\times 10^2$&$4.1\times 10^4$&0.32\\\hline

10&$3\times 10^{10}$&$1$&$10^{-3}$&0&0.1&
$3.7\times 10^2$&$3.5\times 10^2$&$1.2\times 10^2$&$8.7\times 10^2$&$4.8 \times  10^2$ &$5.6\times 10^3$&$2.4\times 10^2$&$7.5\times 10^4$&0.29\\\hline

11&$3\times 10^{10}$&$1$&$10^{-2}$&0&1&
$3.6\times 10^3$&$3.6\times 10^3$&$1.6\times 10^3$&$1.2\times 10^4$&$5.0 \times  10^3$ &$5.4\times 10^4$&$2.4\times 10^2$&$3.1\times 10^4$&0.19\\\hline

12&$3\times 10^{10}$&$1$&$10^{-1}$&0&1&
$6.7\times 10^5$&$7.1\times 10^3$&$1.4\times 10^3$&$1.5\times 10^4$&$9.2 \times  10^3$ &$5.1\times 10^4$&-&-&0.19\\\hline

13&$3\times 10^{10}$&$1$&$10^{-3}$&2.35&1&
$2.7\times 10^3$&$2.7\times 10^3$&$2.2\times 10^3$&$4.4\times 10^4$&$6.4 \times  10^3$ &$1.8\times 10^5$&$2.4\times 10^2$&$4.1\times 10^4$&0.073\\\hline

14&$10^{10}$&$1$&$10^{-3}$&0&1&
$4.1\times 10^3$&$4.1\times 10^3$&$2.7\times 10^3$&$2.3\times 10^4$&$5.1 \times  10^3$ &$7.6\times 10^4$&$2.4\times 10^2$&$5.9\times 10^4$&0.076\\\hline

15&$10^{10}$&$1$&$10^{-2}$&0&1&
$4.1\times 10^3$&$4.1\times 10^3$&$2.7\times 10^3$&$2.3\times 10^4$&$5.1 \times  10^3$ &$7.6\times 10^4$&$2.4\times 10^2$&$5.9\times 10^4$&0.076\\\hline

16&$10^{10}$&$1$&$10^{-1}$&0&1&
$4.1\times 10^3$&$4.0\times 10^3$&$2.6\times 10^3$&$2.2\times 10^4$&$5.0 \times  10^3$ &$7.4\times 10^4$&$2.4\times 10^2$&$5.7\times 10^4$&0.076\\\hline

17 & $10^{10}$ & $1$ & $1$ & 0 &1&
$6.6\times 10^5$&$1.3\times 10^3$&$3.8\times 10^2$&$1.1\times 10^4$&$1.5 \times  10^3$ &$1.3\times 10^4$&-&-&0.076\\\hline

18 & $10^{9}$ & $1$ & $10^{-3}$ & 0 &1&
$1.2\times 10^3$&$1.1\times 10^{3}$&$5.2\times 10^2$&$9.6\times 10^4$&$1.0 \times  10^3$ &$9.7\times 10^3$&$2.4\times 10^2$&$2.4\times 10^5$&0.085\\\hline

19 & $10^{9}$ & $1$&$1$&0&1&
$2.6\times 10^2$&$1.7\times 10^2$&$1.5\times 10^2$&$9.0\times 10^4$&$1.6 \times  10^2$ &$3.2\times 10^3$&$2.4\times 10^2$&$2.5\times 10^5$&0.085\\\hline

20&$10^{8}$&$1$&$10^{-3}$&0&1&
$3.2\times 10^2$&$2.8\times 10^2$&$1.3\times 10^2$&$3.6\times 10^5$&$2.2 \times  10^2$ &$2.5\times 10^3$&$2.4\times 10^2$&$8.8\times 10^5$&0.081\\\hline

21&$10^{7}$&$1$&$10^{-3}$&0&1&
$2.0\times 10^2$&$1.9\times 10^2$&$56$&$1.2\times 10^6$&$1.6\times 10^2$ &$1.1\times 10^3$&-&-&0.082\\\hline

22&$10^{6}$&$1$&$10^{-3}$&0&1&
$22$&21&$2.4$&$1.2\times 10^6$&17 &$34$&-&-&0.079\\\hline

23&$10^{11}$&$1$&$10^{-3}$&0&10&
$6.7\times 10^5$&$8.4\times 10^3$&$3.1\times 10^2$&$1.2\times 10^3$&$1.2 \times  10^3$ &$7.7\times 10^2$&-&-&3.1\\\hline

24&$10^{10}$&$1$&$10^{-3}$&0&10&
$6.6\times 10^5$&$5.2\times 10^4$&$4.8\times 10^3$&$6.4\times 10^3$&$6.3 \times  10^3$ &$1.1\times 10^4$&-&-&0.74\\\hline

25&$10^{9}$&$1$&$10^{-3}$&0&10&
$4.2\times 10^3$&$4.2\times 10^3$&$9.0\times 10^2$&$1.4\times 10^3$&$5.3 \times  10^2$ &$2.7\times 10^3$&$2.5\times 10^2$&$3.2\times 10^4$&0.75\\\hline

26&$10^{11}$&$1$&$10^{-3}$&0&100&
$6.6\times 10^5$&$8.0\times 10^3$&$2.0\times 10^2$&$1.6\times 10^2$&$1.4 \times  10^2$ &$94$&-&-&8.8\\\hline

27&$10^{10}$&$1$&$10^{-3}$&0&100&
$9.0\times 10^3$&$8.8\times 10^3$&$3.0\times 10^2$&$2.1\times 10^2$&$1.5 \times  10^2$ &$83$&$4.3\times 10^3$&$5.4\times 10^4$&3.6\\\hline

28&$10^{9}$&$1$&$10^{-3}$&0&100&
$2.8\times 10^2$&$1.3\times 10^2$&$0$&$14$&$2$ &$0$&$2.7\times 10^2$&$8.1\times 10^4$&3.9\\\hline

\end{tabular}
\end{table*}

\begin{figure}
\begin{center}
\includegraphics[width=85mm]{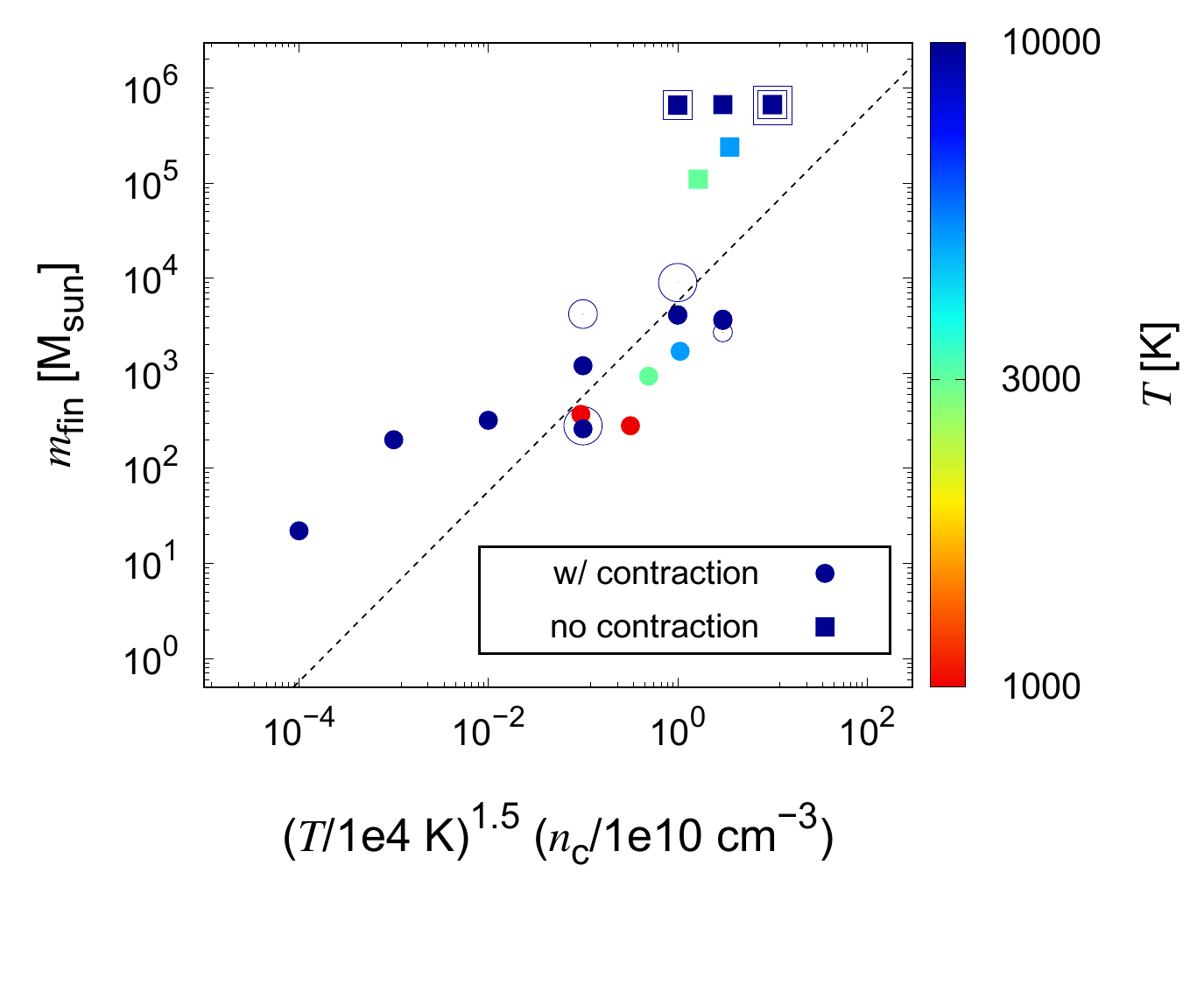}
\caption{
The final mass of the central star as a function of the product $(T/10^4\,\mathrm{K})^{3/2}(n_\mathrm{c}/10^{10}\,\mathrm{cm}^{-3})$. 
Color represents gas temperature, empty circles/squares correspond to models with $\beta=2.35$ (Salpeter mass function), and 
large empty circles/squares correspond to models with high $m_\mathrm{0,max}$. 
The results for cases in which the central star contracts 
and does not contract are shown by circles and squares, respectively. 
The results for cases in which the central star contracts (circles) roughly follow the relation $(m_\mathrm{fin}/5700\,\Msun)=(T/10^4\,\mathrm{K})^{1.5}(n_\mathrm{c}/10^{10}\,\mathrm{cm}^{-3})$ (dashed diagonal line). 
}
\label{fig:mass_n}
\end{center}
\end{figure}

\begin{figure}
\begin{center}
\includegraphics[width=85mm]{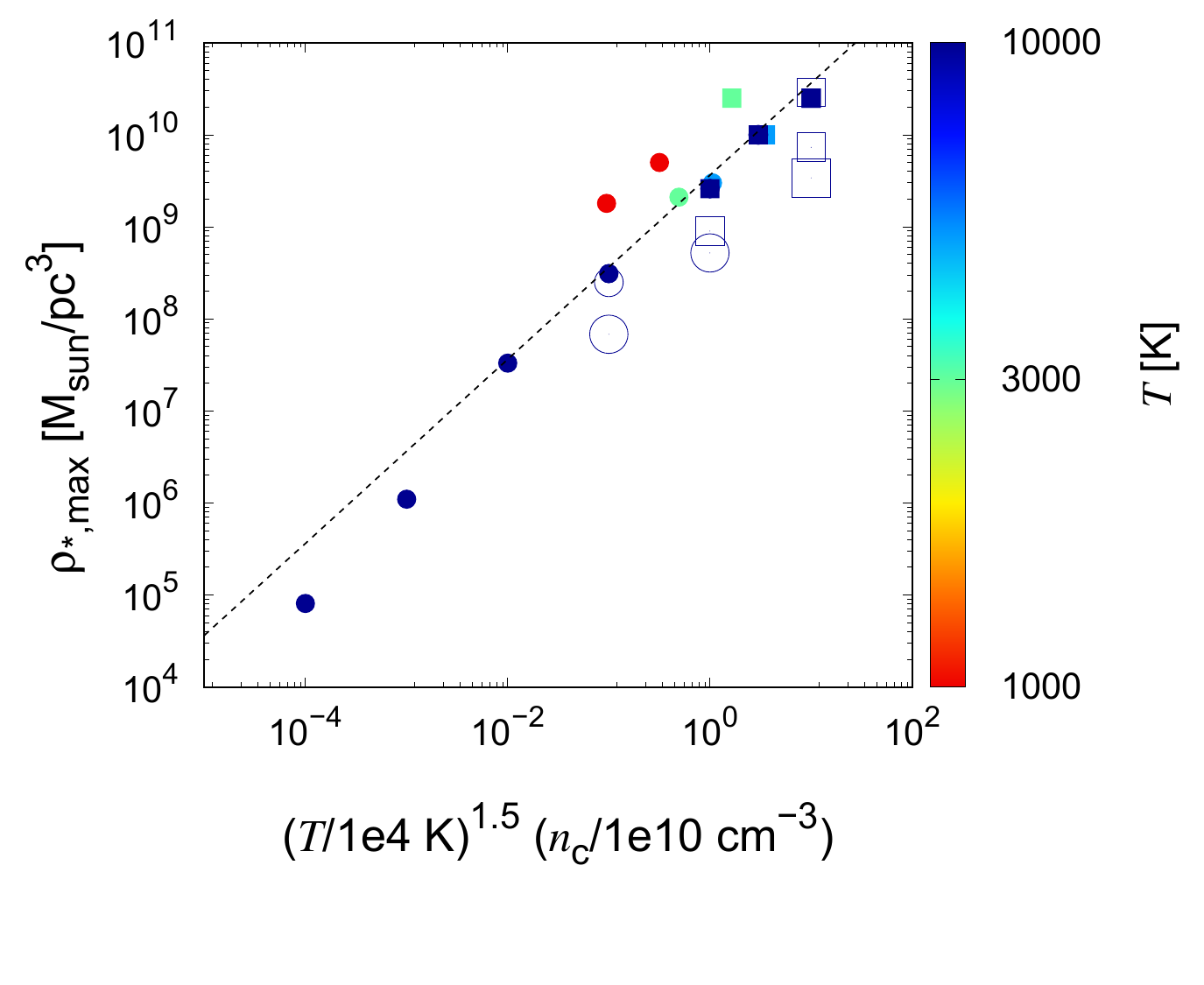}
\caption{
Similar to Fig.~\ref{fig:mass_n}, but the $y$-axis represents the maximum stellar density within the core radius. 
}
\label{fig:mass_ps}
\end{center}
\end{figure}

\subsection{Parameter dependence}
\label{sec:parameter_depend}

The dependence of the results on the input parameters of our model is illustrated through a range of model variants listed in Table~\ref{table_results}. 
The final mass of the central star ($m_\mathrm{fin}$) is most strongly influenced by whether 
the central star contracts or not, 
which in turn depends on the parameters we investigated.
This is illustrated by the masses shown in Figure~\ref{fig:mass_n}.

We find that for efficient growth via stellar bombardment, 
the formation of a high-density star cluster is required in order to enhance the inward acceleration by stellar dynamical friction. In cases with high core gas density $n_\mathrm{c}$, the core radius $r_\mathrm{c}$ is small, and since surrounding stars form at the core radius, the growth rate of the density of stars in early phases is high (equation \ref{eq:rc_m}). 
The growth rate of the stellar density is also high for the high $T$ cases, 
since the star formation rate during the early stages is mostly given by the gas inflow rate (${\dot M}_\mathrm{in}\propto T^{3/2}$, equation~\ref{eq:Mdotin}). 
In high stellar density environments, 
the migration time due to stellar dynamical friction is short and the rate of stellar bombardment is high. 
When the growth rate by stellar bombardment exceeds the critical rate, the central star continues growing without ejecting gas, as seen in the evolution for Model 1 in Figure~\ref{fig:cent_ev_fid}. 
From Table \ref{table_results}, 
for $n_\mathrm{c}= 10^{11}\,\mathrm{cm^{-3}}$ with $T\geq 3\times 10^3$ (Models 1, 3, and 4) 
the stellar accretion rate onto the central star exceeds the critical rate for contraction before $t_\mathrm{KH,ZAMS}$ for the central star, allowing it to grow to $\gtrsim 10^5\,\Msun$.

Whether the central star contracts is also influenced by the value of $f_\mathrm{red}$ (see \S~\ref{sec:stellar_evolution} and \S~\ref{sec:accretion}).
This is the factor by which the gas accretion rate is assumed to be reduced by both fragmentation and by the removal of gas that is captured by the fragmented clumps, 
when the Bondi mass becomes larger than the star's own mass. A high $f_\mathrm{red}$ value increases the gas accretion rate for $M_\mathrm{Bondi}>m_\mathrm{cent}$, which is satisfied for $m_\mathrm{cent}\gtrsim 30\,\Msun$ in Models~1~and~2 (blue and black lines in panel (a) of Figs.~\ref{fig:cent_ev_fid}~and~\ref{fig:cent_ev_cont}). So if the central star can grow to $m_\mathrm{cent}\gtrsim 30\,\Msun$ within $t_\mathrm{KH,ZAMS}$, the high value for $f_\mathrm{red}$ can aid to enhance the growth rate of the central star. 
From Table~\ref{table_results}, we see that
for $n_\mathrm{c}= 3\times 10^{10}\,\mathrm{cm^{-3}}$ with $f_\mathrm{red}\geq 0.1$ (Model 12) 
or $n_\mathrm{c}= 10^{10}\,\mathrm{cm^{-3}}$ with $f_\mathrm{red}\sim 1$ (Model 17), 
the central star keeps expanding 
until 3~Myr when the SMS collapses to a massive BH or when any of the surrounding stars explode as a supernova and blow away all of the gas from the vicinity. 
In Models 12 and 17, the central star grows mainly via stellar accretion until $m_\mathrm{cent}\sim 100\,\Msun$, and then the gas accretion rate onto the central star exceeds the critical rate ${\dot m_\mathrm{cri}}$. 
Thus even for high $f_\mathrm{red}$, stellar accretion is important to enhance the gas accretion rate. 
Unfortunately, the relevant value for $f_\mathrm{red}$ is highly uncertain. To assess it, we need to consider  fragmentation of gas inside the Bondi radius, and the evolution of any accretion disk around the central star. These issues are beyond the scope of the present paper and will be investigated elsewhere in the future. 

In those cases in which the central star contracts and gas is ejected before 3 Myr ($t_\mathrm{ej}<$3~Myr), gas accretion contributes very little to the final mass (see the values of $m_\mathrm{fin}$ and $m_\mathrm{acc,*}$ in Table \ref{table_results}). 
In these cases, since the growth rate should correlate with the efficiency of stellar dynamical friction, which depends on the stellar density, the final mass of the central star should correlate with the stellar density. 
We further assume that 
the stellar density is proportional to the star formation rate over the core radius cubed ($\rho_\mathrm{*}\propto \sim {\dot m}_\mathrm{SF}/r_\mathrm{c}^3$). Due to the scaling relations ${\dot m}_\mathrm{SF}\propto \sim T^{3/2}$ and $r_\mathrm{c} \propto n_\mathrm{c} ^{-1/3}$ (equation~\ref{eq:rc_m}), 
we can expect $\rho_\mathrm{*}\propto \sim T^{3/2} n_\mathrm{c}$. 
Figure~\ref{fig:mass_n} shows the relation between the final mass for the central star and the product $T^{3/2} n_\mathrm{c}$. 
We can indeed see the rough correlation between the final mass and $T^{3/2} n_\mathrm{c}$ as expected in the cases in which the central star contracts (dashed line and circles in Figure~\ref{fig:mass_n}). 
Also, the expected relation $\rho_\mathrm{*}\propto \sim T^{3/2} n_\mathrm{c}$ is roughly confirmed in Fig.~\ref{fig:mass_ps}, which shows the maximum stellar density within the core radius as a function of $T^{3/2} n_\mathrm{c}$. 
The offsets for high $m_\mathrm{max,0}$ or low $T$ cases (large empty or red circles/squares in Fig.~\ref{fig:mass_ps}) are presumably due to the differences in the efficiency of migration, which changes the accretion rate of surrounding stars onto the central star and so affects the stellar density within the core radius.

As discussed above, if $T$ is low, $\rho_*$ remains low, and therefore stellar dynamical friction is inefficient. 
Additionally, since the mass of the stellar cluster in the core is approximately limited by $\dot{M}_{\mathrm{in}}t$, if $T$ is low, then $\dot{M}_\mathrm{in}$ is low (equation~\ref{eq:Mdotin}) and the cluster mass grows slowly. 
If the cluster mass remains low, the number of surrounding stars that bombard the central star is reduced. 
This is plausibly the reason why the final mass of the central star at some fixed values of the combination $T^{3/2}n_\mathrm{c}$ in low $T$ models is lower than those for high $T$ models (Figure~\ref{fig:mass_n}). 
Also in those cases when the central star keeps expanding until it collapses into a massive BH,
the growth rate is determined primarily by the gas accretion rate in the final phase, the final mass depends on the inflow rate and accordingly the gas temperature (square plots in figure \ref{fig:mass_n}). 
This dependence explains why the final masses in Models 1, 6, and 7, which have the same temperature, are the same. 
Note that, in the cases in which the central star keeps expanding for 3~Myr, 
almost all of the gas that fell in from large scales is converted to the central star ($m_\mathrm{fin}\sim 3\,\mathrm{Myr}\times {\dot M}_\mathrm{in}$).

On the other hand, the power-law slope $\beta$ of the IMF
has only a small effect on the final mass (empty circle and square in Figure~\ref{fig:mass_n}).  This is because $\beta$ has almost no effect on the density and mass of the stellar cluster, which are the critical factors for the efficiency of migration by stellar dynamical friction.

For high $m_\mathrm{0,max}$, the final mass is higher than that for low $m_\mathrm{0,max}$ except in Model 28 (large empty symbols in Figure~\ref{fig:mass_n}). When the masses of surrounding stars are high, stellar dynamical friction operates prominently, which facilitates the growth of the central star. For high-mass stars, gas dynamical friction also operates efficiently, which enhances the heating of gas. 
In our models, 
when $\Gamma_\mathrm{GDF}>\Lambda_\mathrm{dust}$, star formation is assumed to stop operating due to the gas heating. In Model 28, star formation becomes inefficient due to this effect, resulting in a low final mass of the central star. 
However, our models cannot predict the evolution in this case, since the gas distribution and accretion processes will be affected by the increased gas pressure. 
Additional studies using three-dimensional hydrodynamical simulations are required to estimate the evolution in these cases, i.e. when $\Gamma_\mathrm{GDF}\gtrsim\Lambda_\mathrm{dust}$.

\section{Discussion}

In this section, we discuss assumptions in our models. 

\subsection{Inconsistency between assumptions}

To calculate the acceleration rate due to stellar dynamical friction, the velocity dispersion of surrounding stars is assumed to be isotropic. 
Such a thermalized distribution for the surrounding stars is realized during the evolution due to the nonresonant and resonant relaxation processes \citep{Kocsis11}. 
We note that, technically, this isotropic distribution is inconsistent with the assumption that the surrounding stars follow circular orbits (in the calculation of the migration rate in equation~\ref{eq:migration_rate}).
However, since migration rates for nonzero- and zero-eccentricity surrounding stars are the same 
when the binding energy is dissipated by the same amount, this inconsistency should have a negligible
impact on our results (i.e. on the evolution of the central star). 

There is an inconsistency between the star-formation prescription, assuming surrounding stars form in a rotating gas disk,  and equation (\ref{eq:sdf}), which assumes that surrounding stars are isotropically distributed. We expect that this does not significantly affect our conclusions. First, the gas disk thickness $h/r$ roughly evolves from $\sim0.4$ to $\sim0.08$ from $10^3$ yr to $10^5$ yr in Model 1, and never reaches very small values. Second, we expect that an isotropic distribution is established by relaxation processes \citep[e.g.][]{Kocsis11}. Finally, even if relaxation processes are inefficient, stellar dynamical friction would operate more strongly in a disk configuration, due to the higher stellar density and the low relative velocity between surrounding stars, which would facilitate stellar accretion. Thus the isotropic distribution of surrounding stars is a conservative choice for the growth rate of the central star.

Although the disk around the central star is thick, we used the approximation $h\sim c_s/\Omega$ for the scale height of the disk. 
At $h/r=0.4$, the Toomre $Q$ parameter is overestimated by $\sim 10\%$ due to the approximation. Since $Q$ depends linearly on $n_\mathrm{c}$, this assumption may affect the dependence of the results on $n_\mathrm{c}$ by the same factor of 10\%, which is well within other uncertainties of our simplified model.

\subsection{Star formation efficiency}

In our models, we allow a high star formation efficiency (SFE), defined as the ratio of the total mass in newly formed stars to the initial gas mass. 
For example, the SFE within the core radius is $\sim 0.7$ at $t=10^4$ yr in our fiducial Model 1 (see below), and it increases with time. Observationally, some massive molecular clouds are found to have an SFE of $>0.5$ \citep{Turner15}, though the SFEs of most molecular clouds in the Milky Way are $\sim 0.002-0.3$ \citep{Murray11}. 
On the other hand, theoretically the SFE is determined by radiation pressure from ionizing ultraviolet (UV) photons, nonionizing UV photons, and infrared (IR) photons \citep[e.g.][]{Kim18}. Radiation pressure from nonionizing UV photons does not halt gas collapse when the gas surface density exceeds a critical value \citep{Raskutti16,Thompson16}, and likewise IR photons do not halt collapse unless the IR opacity is very high \citep{Skinner15}. 
In our models, the gas surface density within the core radius (Eq.~\ref{eq:ngas}) is much higher than the critical value \citep{Raskutti16,Thompson16}, and the IR opacity is extremely low because the gas is metal poor.

We also estimate whether ionizing UV photons are confined within the Bondi-Hoyle-Lyttleton radius of each star (in which case they do not halt gas collapse; \S~\ref{sec:gas_ejection}). 
According to numerical simulations \citep{Skinner15,Raskutti16,Thompson16,Kim18}, when these feedback effects are inefficient, the SFE is  close to unity (but not exactly 1 in their simulations due to the initial turbulent motion). Thus we considered the SFE of $\sim1$  to be justified in our case. 
On the other hand, although the SFE within the core radius becomes close to 1, 
the SFE within the rest of the halo is still low in our models since the baryon mass within the halo is $\sim 2\times 10^6\,\Msun$, and the mass of the stellar cluster is at most $\sim 10^4\,\Msun$ (see orange line in panel (a) of Figure \ref{fig:cent_ev_fid} below), so the SFE might not be so extreme compared to the SFE observed in  molecular clouds (0.002-0.5). Also as mentioned earlier, the rate of star formation is sensitive to the gas temperature in our models. Compared to the fiducial case of $T=10^4$ K, the star formation rates for $T=5\times 10^3, 3\times 10^3,$ and $10^3$ K are lower by factor of 2.8, 6.1, and 32, respectively. These lower-$T$ models may be considered as proxies for lower SFE cases. 

\subsection{The eccentric orbit}

In this study, surrounding stars are allowed to migrate in- or outward, but are assumed to remain on circular orbits. If angular momentum exchange dominates the accretion of surrounding stars, stellar accretion becomes more efficient than in our model, since the binding energy of a surrounding star required to accrete onto the central star decreases by a factor of $1/(1-e_i)$ where $e_i$ is the eccentricity of the $i^{\rm th}$ surrounding star. 
To investigate the impact of nonzero eccentricity, 
we examine a case in which the eccentricity distribution for surrounding stars is assumed to be thermalized (e.g. due to two-body relaxation), and has a distribution function of $f(e_i)=2e_i de_i$~\citep[e.g.][]{Jeans1919,Heggie75}. 
In this case, the central star captures surrounding stars from the larger distance $r_i=R_\mathrm{cent}/(1-e_i)$ (this is the only difference from the models above). Simulating this prescription with the parameter set of Model 2, 
we find the final mass of the central star to be $m_\mathrm{fin}=4.6\times 10^3\,\Msun$, which is almost unchanged from the final mass in Model 2  ($m_\mathrm{fin}=3.7\times 10^{3}\,\Msun$). 
However, this neglects other possible effects. For example, a surrounding star with a high eccentricity interacts with stars and gas orbiting over a wider ranges of $r$,
and mass loss should increase when a surrounding star with extremely  high eccentricity is captured.  

If most stellar accretion onto the central star is highly eccentric, and the mass lost at stellar accretion is typically a large fraction of the mass of the accreted surrounding star, the results of our models may be largely influenced. 
We intend to explore these issues in a follow-up study, based on direct $N$-body and hydrodynamical simulations.
Here we only briefly consider the possible fate of the lost gas.   
If the launch velocity of this gas is similar in magnitude to the collision velocity between the stars, 
then the gas is kicked out to at most the apocenter of the colliding star's orbit before the collision. 
On the other hand, due to the low specific angular momentum of the ejected gas, it would be circularized (presumably by shocks it encounters) near the central star, similar to the expectation in the context of tidal disruption of stars \citep{Hayasaki2013,Hayasaki2016,Bonnerot_Lu2019}. 
In the vicinity of the central star, 
the viscous timescale is very short. 
Thus the gas ejected in high-eccentricity collisions may end up promptly accreted onto the central star, leaving 
our results largely unchanged.

\subsection{Evolution following the formation of the massive black hole}

Finally, let us consider the evolution of the stellar cluster after the central star collapses to a massive BH. 
Since collisions, relaxation, and evaporation are important mechanisms for cluster evolution, 
we show the collision (black curve in the bottom panel of Figure \ref{fig:snapshot}), stellar dynamical friction (orange curve), and evaporation (red curve) timescales for the stellar cluster at 3~Myr in Model 2. 
For the collision timescale (equation \ref{eq:n_coll}), the collision radius is assumed to be twice the radius of stars with the average mass, and stars are assumed to be in the contracted phase (equation \ref{eq:r_contract}). 
We adopt the evaporation timescale to be $t_{\mathrm{evap},l}=f_\mathrm{evap} t_{\mathrm{relax},l}$ \citep{Binney08}, 
where the factor $f_\mathrm{evap}$ is $\sim 300$ for clusters with a single stellar mass, and without a massive black hole and gas \citep{Spitzer87}, $t_{\mathrm{relax},l}=0.34 \sigma_{*,l}^3/(G^2 \overline{m}_l \rho_{*,l} \mathrm{ln}\Lambda)$ is the relaxation timescale \citep{Binney08}, and we set the Coulomb logarithm to be 10. 
Although we set $f_\mathrm{evap}= 300$, this value may be significantly increased for the cluster with a central massive BH which may help to retain objects from dynamical ejections both by increasing the cluster's escape velocity and by inhibiting binary formation.\footnote{The binary formation rate due to three-body encounters scales with $\sigma_{*}^{-9}$, which is greatly affected by a massive black hole \citep{Binney08}.}
Figure~\ref{fig:snapshot} shows that the collision and  evaporation timescales in the outer regions of the cluster are longer (at $\gsim 2000$ AU where $\rho_* \sim 10^{7-8}\,\Msun/\mathrm{pc}^3$) and comparable to the Hubble time of $\sim 10\,\mathrm{Gyr}$, respectively. 
Thus these clusters could possibly survive to low-$z$ epochs. 

If such high-density clusters sink to the centers of massive local galaxies, the relics of such high-density clusters formed at high $z$ may be observationally confused with stellar systems formed at lower redshift, such as infalling dense clusters and in situ formed stars, if those produce similarly high stellar density environments. 
The stellar density of nuclear star clusters may also be reduced by a supermassive black hole binary following galaxy collisions \citep{Merritt2006}. 
On the other hand, if such clusters remain isolated, their relics may in principle be clearly identified in the local universe. 
Such clusters contain low-mass and extremely low-metallicity stars, and an intermediate-mass BH with the mass of $\sim 10^3 \,\Msun$. 
Stellar densities within $\sim 2000$ AU of galactic nuclei have not been resolved to date \citep[e.g.][]{Nguyen18}. 
Extrapolating the observed density profile from diffuse light in the center of the Milky Way, the stellar mass within $\sim$2000 AU from Sgr A* is estimated to be $\sim 600-800\,\Msun$ \citep[e.g.][]{Schodel18}, which is about 
a factor $\sim 3$
smaller than that for high-density clusters formed at high $z$ (middle panel of figure \ref{fig:snapshot}). 
If such high-density nuclear star clusters are identified with low-mass stars in the future, 
they might represent the fossils of high-$z$ clusters.

In the stellar cluster in Model 2, the accretion of stars will continue after the BH formation.  This may contribute to the rate of high-$z$ tidal disruption events \citep{Kashiyama16} or to gravitational wave events observed by 
the {\it Laser Interferometer Space Antenna} ({\it LISA}) \citep[e.g.][]{AmaroSeoane07,Hartwig18}.

\begin{figure}
\begin{center}
\includegraphics[width=90mm]{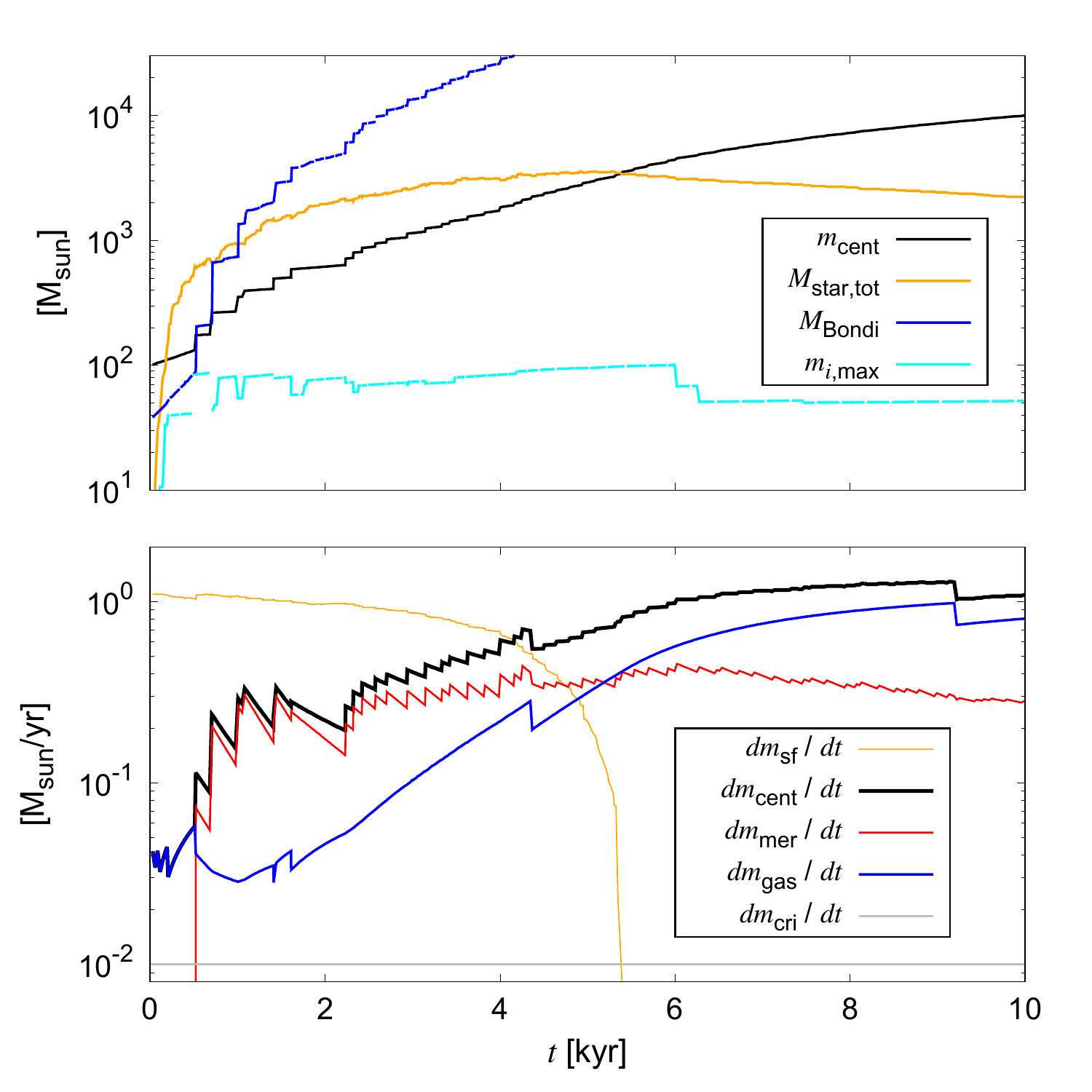}
\caption{
Same as panels (a) and (c) in Fig.~\ref{fig:cent_ev_fid}, but parameter settings are different (see \S \ref{sec:comparison}). 
}
\label{fig:cent_ev_cont2}
\end{center}
\end{figure}

\subsection{Comparison with hydrodynamical simulations}
\label{sec:comparison}

After this work was submitted and posted on {\it arXiv}, \citet{Chon20} presented results for three-dimensional hydrodynamical simulations focusing on similar scenarios. 
In this section, we briefly discuss the consistency between our predictions and their simulation results. 

\citet{Chon20} chose a relatively massive halo formed from cosmological initial conditions, in which the gas inflow rate is very high ($\gtrsim 1\,\Msun/\mathrm{yr}$). 
They find that the power law index of the initial mass function is $\sim -1$, the maximum mass of surrounding stars is $\sim 100\,\Msun$, and stars form at densities $\gtrsim 10^{11}\,\mathrm{cm}^{-3}$. Referring to these findings, we perform models with $n_\mathrm{c}=10^{11}\,\mathrm{cm}^{-3}$, $m_\mathrm{0,max}=100\,\mathrm{Msun}$, $\beta=-1$, and $T=3\times 10^4\,\mathrm{K}$.  
Since they use a barotropic equation of state, we allow the star formation even when $\Gamma_\mathrm{GDF}>\Lambda_\mathrm{dust}$. 
Since $f_\mathrm{red}$ is highly uncertain, 
we varied $f_\mathrm{red}$ between $10^{-3}$ and 1. 

The evolution of $f_\mathrm{red}=0.1$ is shown in Fig.~\ref{fig:cent_ev_cont2}. 
The central star grows to $\sim 10^{4}\,\Msun$ by $10$ kyr (black line in the upper panel) mainly due to mergers in the early phases (red line in the lower panel) and gas accretion in the later phases (blue line in the lower panel). 

The total stellar mass stops increasing at $\sim 5$ kyr for $f_\mathrm{red}=0.1$ (orange line in the upper panel) while it keeps increasing for $f_\mathrm{red}\leq0.01$. 
Similar trends are seen in panel (d) of Fig.~4 in \citet{Chon20}; namely the number of stars keeps increasing for $Z=10^{-4}$ and $Z=10^{-5}$, while it stops increasing at $\sim 5$ kyr for other models. We conclude from our models that quenching of star formation in their simulations is related to whether the accretion rate onto the central star comes close to the gas inflow rate. 

For $f_\mathrm{red}=$0.001, 0.01, 0.1, and 1, respectively, 
the mass of the central star at $10$ kyr is
$5.5\times 10^3$, $5.8\times 10^3$, $1.0\times 10^4$, and 
$1.1\times 10^4\,\Msun$, 
the number of surviving stars is
896, 743, 426, and 203, 
and the contribution of mergers to the total final mass of the central star 
is 98$\%$, 87$\%$, 33$\%$, and 6.8$\%$. 
\citet{Chon20} find that the contribution of mergers is $\sim 30-70\%$, the central mass is $\sim10^{4}\,\Msun$, and the number of surviving stars is $\sim 500-4000$ at $10$ kyr. Thus our models with $f_\mathrm{red}\sim 0.01-0.1$ can reproduce their results remarkably well. 
The larger number of surrounding stars in \citet{Chon20} presumably reflects the lower minimum mass of surrounding stars ($\sim 0.01\,\Msun$) for $Z\sim 10^{-3}-10^{-5}$ compared to the value adopted in our models ($0.08\,\Msun$). 

When we continue the above models beyond the $10$ kyr at which \citet{Chon20} stopped their simulation, we find that the final mass (at $t=3$ Myr) of the central star for $f_\mathrm{red}=10^{-3}-1$ is as high as $3.5\times 10^6$, due to the high inflow rate (${\dot M}_\mathrm{in}\sim 1.1\,\Msun/\mathrm{yr}$).

\section{Conclusions}

In this paper, we propose a process for forming supermassive stars via stellar collisions and accretion in high-redshift protogalaxies. 
The scenario envisioned here shares some aspects of both the popular ``direct collapse'' and the ``runaway collision'' scenarios. 
We focus on environments in which a gas cloud is polluted only by a moderate amount of metals, and its ${\rm H_2}$ abundance is suppressed.
In such environments, a gas cloud fragments only at very high density, producing a high-density stellar cluster~\citep{Omukai08}. 
If gas is ejected soon after stars form, the final mass of a central star becomes $\sim10^3\,\Msun$ \citep{Sakurai17}. 

The novel aspect proposed here is that if subsequent frequent capture and accretion of surrounding stars onto a central star efficiently heats the envelope of the central star, 
the central star continues expanding, and  
gas will be retained in the system due to the lack of strong UV radiation and weak photoionization feedback from the bloated central star.
The central star can therefore keep growing until 
the supply of surrounding stars and gas runs out due to gas ejection by 
SN explosions or by accretion feedback from a collapsed massive BH.
We call such a rapid stellar accretion process ``stellar bombardment'', which 
could be caused by efficient stellar migration via relaxation processes, the increase of the stellar radius by the mass increase, and most importantly, the heating and bloating of the stellar envelope due to the frequent stellar accretion itself.

To investigate the viability of this ``stellar bombardment'' scenario, we have performed numerical modeling using a semianalytic toy model. The model includes dynamical friction by stars and gas, star formation, gas accretion, collisions, and gas ejection. Our main results can be summarized as follows: 

\begin{enumerate}
\item 
When the central core density exceeds $10^{11}\,\mathrm{cm^{-3}}$ and the gas temperature is $\geq 3\times 10^3$ K, 
the central star continues growing 
without contracting until it reaches a mass of $\sim 10^5-10^6\,{\rm M_\odot}$ at 3~Myr.
The central star grows mainly by stellar bombardment early on,  and by gas accretion in the later phases. 

\item
When the central core density is below $3\times 10^{10}\,\mathrm{cm^{-3}}$, the central star contracts due to the subcritical rate of accretion and heating by surrounding stars. After the contraction, photoionization feedback ejects gas from the system, reducing the final central star mass by about two orders of magnitude, to $\lesssim10^4\,{\rm M_\odot}$. 

\item
The final mass of the central star depends strongly on the gas temperature and the core density of the gas, in addition to whether the central star contracts (figure \ref{fig:mass_n}). 
This is because the efficient growth of the central star by stellar accretion requires a high-density cluster. 
High-density star clusters can be realized for
high star formation rates and/or compact core sizes, which in turn are produced for high gas temperature and core gas density, respectively.   In a cosmological setting, these conditions can arise in metal-poor atomic-cooling halos, in which the ${\rm H_2}$ abundance has been suppressed, leading to inefficient cooling until very high densities are reached.

\end{enumerate}

In this paper we have used a simple toy model to illustrate the possibility of this new evolutionary process. 
To understand this pathway in more detail, including its viability, future $N$-body and hydrodynamical simulations will be required, which are able to follow stellar evolution 
and radiation feedback onto the collapsing cloud. 
\citet{Chon20} have recently presented results from three-dimensional hydrodynamical simulations, albeit with a prescribed barotropic equation of state, and without radiation. They find that supermassive stars likely form from gas with  metallicities up to $\sim 10^{-4}\,{\rm Z_\odot}$ due to stellar accretion and gas accretion. Thus our predictions are confirmed by more realistic numerical simulations.

\acknowledgments

We thank to Simon Portegies Zwart and Takashi Hosokawa 
for important suggestions.
This work received funding from the European Research Council (ERC) under the European Union’s Horizon 2020 Programme for Research and Innovation ERC-2014-STG under grant agreement No. 638435(GalNUC), and from the Hungarian National Research, Development, and Innovation Office under grant NKFIH KH-125675.
 ZH acknowledges support from NASA grant NNX15AB19G and NSF grant 1715661.
Simulations and analyses were carried out on Cray XC50 and computers 
at the Center for Computational Astrophysics, National Astronomical Observatory of Japan.

\bibliographystyle{yahapj.bst}
\bibliography{runaway}

\end{document}